\documentclass[
pra,aps,
reprint,
a4paper,
amsmath,amssymb,amsbsy,
floatfix
]{revtex4-1}
\usepackage{etoolbox}
	\newtoggle{arXiv}\toggletrue{arXiv}					
\usepackage{newtxtext, newtxmath}
\usepackage{graphicx}
\usepackage{color, calc}
\usepackage{array, booktabs}
	\newcolumntype{L}{>{$}l<{$}}
	\newcolumntype{C}{>{$}c<{$}}
	\newcolumntype{R}{>{$}r<{$}}
\usepackage{tikz, tikz-3dplot}
	\usetikzlibrary{arrows, arrows.meta, calc, positioning}
	\usetikzlibrary{decorations.pathmorphing}	
	\tikzset{>=Stealth}
\usepackage{pgfplots}
	\pgfplotsset{compat=1.14, small}
\usepackage{siunitx, physics}
	\newcommand{\I}{\ensuremath{\mathrm{i}}}
	\newcommand{\Exp}[1]{\mathrm{e}^{#1}}
\usepackage{enumitem}
\setlist[description]{labelindent=0pt, leftmargin=\parindent, font=\normalfont\itshape}
\usepackage{multirow}
\usepackage{xfrac}
\usepackage{hyperref}

\newcommand{\av}[1]{\ensuremath{\langle #1 \rangle}}
\newcommand{\rbra}[1]{\left(#1\right)}
\newcommand{\sbra}[1]{\left[#1\right]}

\begin{document}


\title{Probing the limits of correlations in an indivisible quantum system}

\author{M. Malinowski}\email{maciejm@phys.ethz.ch}
\author{C. Zhang}
\author{F. M. Leupold}
\author{J. Alonso}\email{alonso@phys.ethz.ch}
\author{J. P. Home}\email{jhome@phys.ethz.ch}
\affiliation{Institute for Quantum Electronics, ETH Z\"urich, Otto-Stern-Weg 1, 8093 Z\"urich, Switzerland}
\author{A. Cabello}
\affiliation{Departamento de F\'isica Aplicada II, Universidad de Sevilla, 41012 Sevilla, Spain}


\begin{abstract}
	
We employ a trapped ion to study quantum contextual correlations in a single qutrit using the 5-observable KCBS inequality, which is arguably the most fundamental non-contextuality inequality for testing Quantum Mechanics (QM). We quantify the effect of systematics in our experiment by purposely scanning the degree of signaling between measurements, which allows us to place realistic bounds on the non-classicality of the observed correlations. Our results violate the classical bound for this experiment by up to 25 standard deviations, while being in agreement with the QM limit. In order to test the prediction of QM that the contextual fraction increases with the number of observables, we gradually increase the complexity of our measurements from 5 up to 121 observables. We find stronger-than-classical correlations in all prepared scenarios up to 101 observables, beyond which experimental imperfections blur the quantum-classical divide.

\end{abstract}

\maketitle





Quantum contextuality speaks against the classical perception that the act of measurement merely reveals pre-existing, context-independent properties of the measured system. This leads to correlations between observables which are stronger than those in classical physics \cite{67Kochen}. One way to test these is to construct Non-Contextuality (NC) inequalities which are constrained classically but violated by quantum systems. In Ref.~\cite{08Klyachko}, Klyachko, Can, Binicio{\u{g}}lu, and Shumovsky (KCBS) provided an inequality with the lowest possible number of measurement settings, while featuring the largest possible gap between quantum and classical predictions \iftoggle{arXiv}{(App.~\ref{sec:exclusivity}, \cite{14Cabello})}{\cite{14Cabello,SuppMat}}. This requires measuring $N = 5$ observables in a three-level system or qutrit, the smallest quantum state space in which such correlations can be observed. In one extra dimension, the Bell inequality on two qubits also provides a test of non-classical correlations, but with the additional requirement that the qubits are space-like separated \cite{83Heywood,10Cabello}. In this scenario, the observation of non-classical correlations causes the exclusion of locality or of realism.

While tests have been performed which aim to saturate the Tsirelson quantum bound of the Bell inequality \cite{15Poh,15Christensen}, their legitimacy has been recently challenged based on signaling \cite{18Smania}, and there is no consensus on the validity of equivalent previous experimental tests of KCBS \cite{11Lapkiewicz,*13Ahrens,*13Lapkiewicz,*13Deng,SuppMat}. The significance of NC tests is compromised when the assumptions of the underlying theory are not fulfilled in experiments. In particular, the characterization of systematic signaling is notoriously scarce in experimental papers aiming at saturating quantum correlations \cite{18Smania}, despite ongoing theoretical efforts \cite{05Spekkens,*17Kunjwal,15Kujala}. Given that NC inequalities can show a trade-off between signaling and the amount of violation, thoroughly accounting for the former is a major pending task. As a consequence, there is no undisputed experimental evidence to date that the maximum predicted by QM for any NC inequality can be reached \iftoggle{arXiv}{(Apps.~\ref{sec:KCBScomparison}-\ref{sec:Bellcomparison}, \cite{18Smania})}{\cite{18Smania,SuppMat}}.

Bell and KCBS inequalities differ in the size of Hilbert space and the number of observables. Beyond the minimal instances, both the Bell and KCBS inequalities have been extended to larger numbers of observables \cite{13Araujo}. This can be used to constrain the predictive power of beyond-quantum theories \cite{11Colbeck}, to certify the randomness of numbers \cite{13Dhara}, or to quantify the computational power available in a quantum system \cite{17Abramsky}. Extended Bell inequalities (or Bell chained inequalities) have been violated in photonic systems and with trapped ions \cite{15Christensen,17Tan}, with the former showing a violation up to $N = 90$ observables. Nevertheless, current studies on the more fundamental extended KCBS scenario, where even stronger correlations are possible, are limited to $N\leq7$ \cite{15Arias}.

In this Letter, we present experimental results which reach the QM bound of the KCBS inequality, and which exhibit correlations beyond those accessible in a Bell experiment \iftoggle{arXiv}{(App.~\ref{sec:exclusivity}, \cite{13Cabello})}{\cite{SuppMat,13Cabello}}. We also extend the KCBS test to $5 \leq N \leq 121$ and measure stronger-than-classical correlations up to $N = 101$ observables, with the largest contextual fraction $\text{CF} = 0.800(4)$ for $N = 31$ \cite{11Abramsky}. We examine the assumptions of compatible measurements \cite{10Guhne,*14Larsson,18Smania} by purposely scanning the degree of signaling in the experiment, thereby evaluating the trade-off between signaling and violation in NC inequalities. This allows us to quantify and minimize systematic effects, which we use to penalize our results in line with recent theoretical proposals \cite{15Kujala}. In these experiments we combine high-fidelity unitary operations with high-precision projective detection \cite{17Leupold,*16Alonso} to close both the individual-existence loophole (by performing sequential, rather than simultaneous, measurements \cite{10Guhne,16Jerger}) and the detection loophole.




\begin{figure}
	{
		\includegraphics[width= 0.4 \columnwidth]{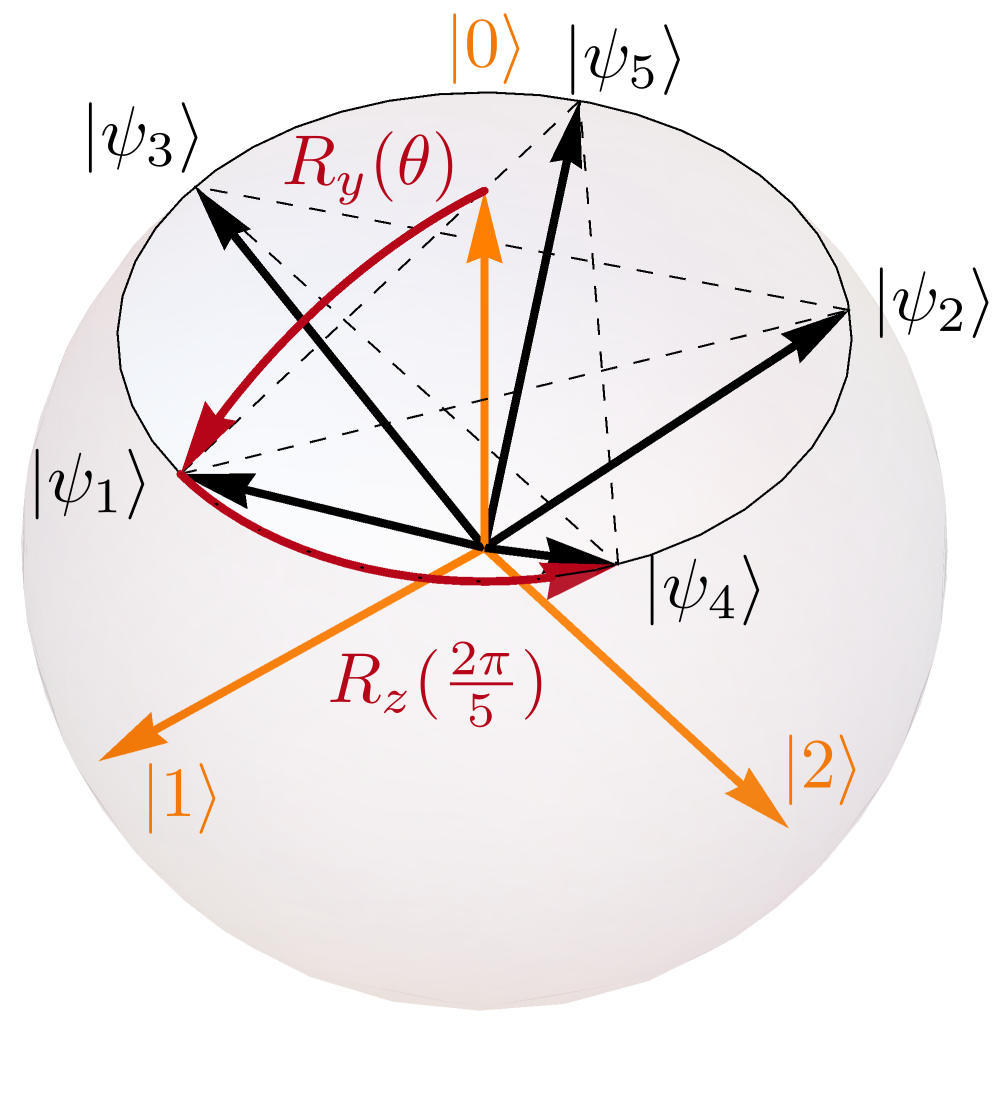}
	}
	{
		\includegraphics[width= 0.57 \columnwidth]{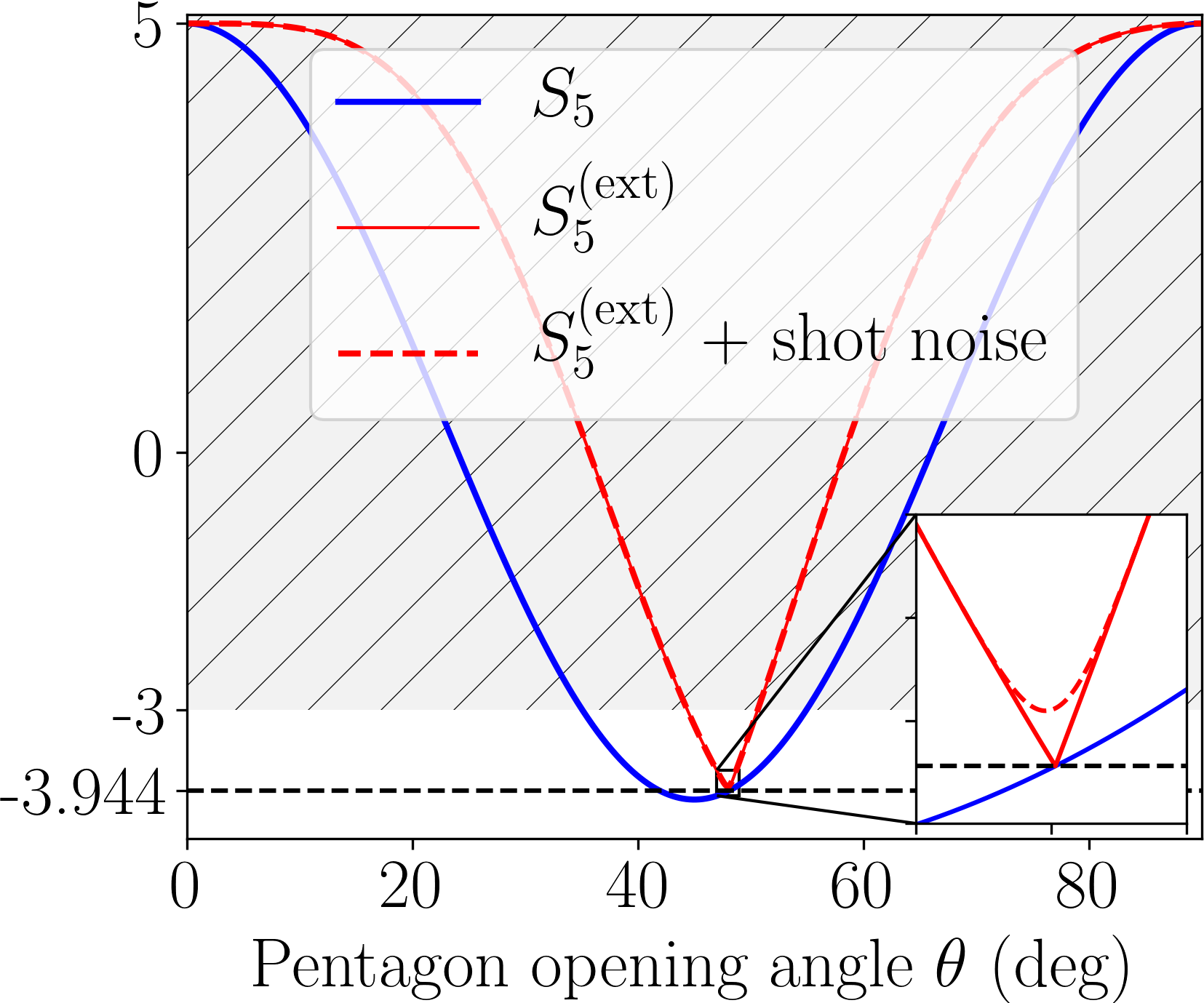}
	}
	
	\caption{(Left): ``Qutrit sphere'' spanned by arbitrary superpositions of the type $\alpha\ket{0} + \beta\ket{1} + \gamma\ket{2}$, where $\alpha,\beta,\gamma\in\mathbb{R}$ and such that $\alpha^2+\beta^2+\gamma^2=1$. The directions along which projective measurements are performed are indicated by states $\ket{\psi_i}$.  When $\theta=\theta_5$, states connected by dashed lines are orthogonal. Experimentally relevant rotations are shown in dark red (see text). (Right): Predicted values of $S_5$ (Eq. (\ref{eq:KCBS})) and $S_5^\text{(ext)}$ (Eq. (\ref{eq:extKCBS})) as a function of $\theta$. The hashed region above $-3$ shows the space where $S_5^\text{(ext)}$ is consistent with NC models. The inset shows how a finite number of experiments (here $5\times 10^4$ per data point) leads to a necessary deviation from theory around $\theta_5$ (dashed line). The analytical formulas for the plotted curves are given in \iftoggle{arXiv}{App.~\ref{sec:theory}}{\cite{SuppMat}}.}
	\label{fig:pentagon}
\end{figure}

We encode the qutrit basis states $\ket{0},\ket{1},\ket{2}$ onto internal electronic energy levels of a single $^{40}$Ca$^+$ ion confined in a cryogenic surface-electrode radio-frequency trap. Combinations of coherent laser pulses resonant with $\ket{0} \leftrightarrow \ket{1}$ and $\ket{0} \leftrightarrow \ket{2}$ transitions allow us to generate arbitrary single-qutrit rotations. Quantum non-demolition measurements of arbitrary observables are achieved with high fidelity by combining these coherent rotations with discrimination of $\ket{0}$ from $\ket{1}$ or $\ket{2}$ using state-dependent fluorescence, allowing the study of correlations between sequential measurements \cite{17Leupold}. Further details of the experimental setup can be found in Refs.~\cite{17Leupold}.

In a first experiment we perform sequences of pairs of measurements along rays defined by a set of $N=5$ states $\ket{\psi_i}$ on a ``qutrit sphere'' of real superpositions of the basis states (see Fig.~\ref{fig:pentagon}). Explicitly, the \textit{pentagon states} are given by
\begin{align}	\label{eq:psis}
\ket{\psi_i} = U_i \ket{0}	\qq{with}
U_i= R_z^{2i-2}( \tfrac{2\pi}{5}) R_y(\theta),
\end{align}
where $R_y(\theta)$ represents a rotation by angle $\theta$ around $\ket{2}$ (associated with the $y$-axis), and $R_z(\frac{2\pi}{5})$ represents a rotation by angle $\frac{2 \pi}{5}$ around $\ket{0}$ (associated with the $z$-axis, Fig.~\ref{fig:pentagon}). Here $i$ is a modulo 5 integer, with $\ket{\psi_0}\equiv\ket{\psi_5}$ and $\ket{\psi_{6}}\equiv\ket{\psi_1}$. We define a convention whereby a sharp measurement \cite{14Chiribella} $M_i$ along ray $\ket{\psi_i}$ will yield one of two values $A_i=\pm1$. If $A_i=1$, the state is projected onto $\ket{\psi_i}$, whereas $A_i=-1$ projects onto an orthogonal space spanned by $\ket{\psi_{i-1}}$ and $\ket{\psi_{i+1}}$, preserving the coherence in that subspace \iftoggle{arXiv}{(App.~\ref{sec:observables})}{\cite{SuppMat}}. When measurements of $M_i$ and $M_j$ are conducted sequentially, with $M_i$ measured first and $M_j$ measured second, we denote their respective outcomes as $A_i^{(1)}$ and $A_j^{(2)}$.

The KCBS scenario, in which measurements $M_i$ and $M_{i\pm 1}$ are compatible (their operators commute), arises when $\theta_5 = \arccos (5^{-1/4}) \approx 48^{\circ}$ and hence $\bra{\psi_{i}}\ket{\psi_{i\pm1}} = 0$. In Fig.~\ref{fig:pentagon}, this is indicated by dashed lines joining pairs of states.  In NC models the sum of correlators is bounded from below:
\begin{equation}
\label{eq:KCBS}
S_5(\theta_5) = S_5^{\pm}(\theta_5)=\sum_{i=1}^{5}\av{A_{i}^{(1)}A_{i\pm1}^{(2)}}\geq \bar{S}_5^\text{NC}=-3.
\end{equation}
This result is called the KCBS inequality \cite{08Klyachko} in either ``normal order'' ($S_5^+$) or ``reverse order'' ($S_5^-$). According to QM:
\begin{equation}
\label{eq:KCBSqm}
S_5(\theta_5)\geq \bar{S}_5^\text{QM}=5-4\sqrt{5}\approx -3.944.
\end{equation}
This is independent of the order and the equality is obtained if and only if the system is initialized to $\ket{0}$ and all experimental operations are carried out with perfect fidelity. Note that mis-calibration of the opening angle $\theta$ may result in $S_5(\theta)<\bar{S}_5^\text{QM}$ (see Fig.\ref{fig:pentagon}). Such results do not reveal non-classical effects, since outcomes of non-compatible measurements can in general be explained by NC models. However, unavoidable experimental imperfections will lead to some degree of signaling and incompatibility. This fact has been identified as the main weakness of contextuality tests in local systems and is often referred to as the ``compatibility'' or ``finite-precision loophole'' \cite{99Kent,*00Clifton,*02Cabello,*16Hu}. The loophole can be addressed by making use of ``extended inequalities'' \cite{10Guhne,15Kujala}. These are modifications of non-contextuality inequalities that do not require perfect compatibility, but can only be used to study either restricted classes of NC models or extended notions of contextuality. Here we follow the latter approach and use the result in Ref.~\cite{15Kujala}, extending the KCBS inequality to
\begin{equation}
\label{eq:extKCBS}
S_5^\text{(ext)}(\theta) =\sum_{i=1}^{5}\av{A_{i}^{(1)}A_{i\pm1}^{(2)}} + \sum_{i=1}^{5}\epsilon_i \geq \bar{S}_5^\text{NC} = -3,
\end{equation}
where $\epsilon_i=\abs{\av{A_{i}^{(1)}}-\av{A_{i}^{(2)}}}$, and thus $\sum_{i=1}^{5} \epsilon_i$ penalizes signaling between measurements. Note that this inequality can be used for both normal and reverse order measurements. $S_5^\text{(ext)}(\theta)$ is plotted in Fig.\ref{fig:pentagon} (red dashed line), showing that a finite range of $\theta$ leads to results inconsistent with NC models (hashed region). Aside from systematic effects, some amount of signaling is expected purely due to shot-noise-limited statistics (otherwise known as quantum projection noise). This means that experimental results are expected to systematically deviate from $S_5^\text{(ext)}(\theta)$ for a finite number of measurements (Fig.~\ref{fig:pentagon}, red dashed line).


\begin{figure}
	\includegraphics[width= 1.0 \columnwidth]{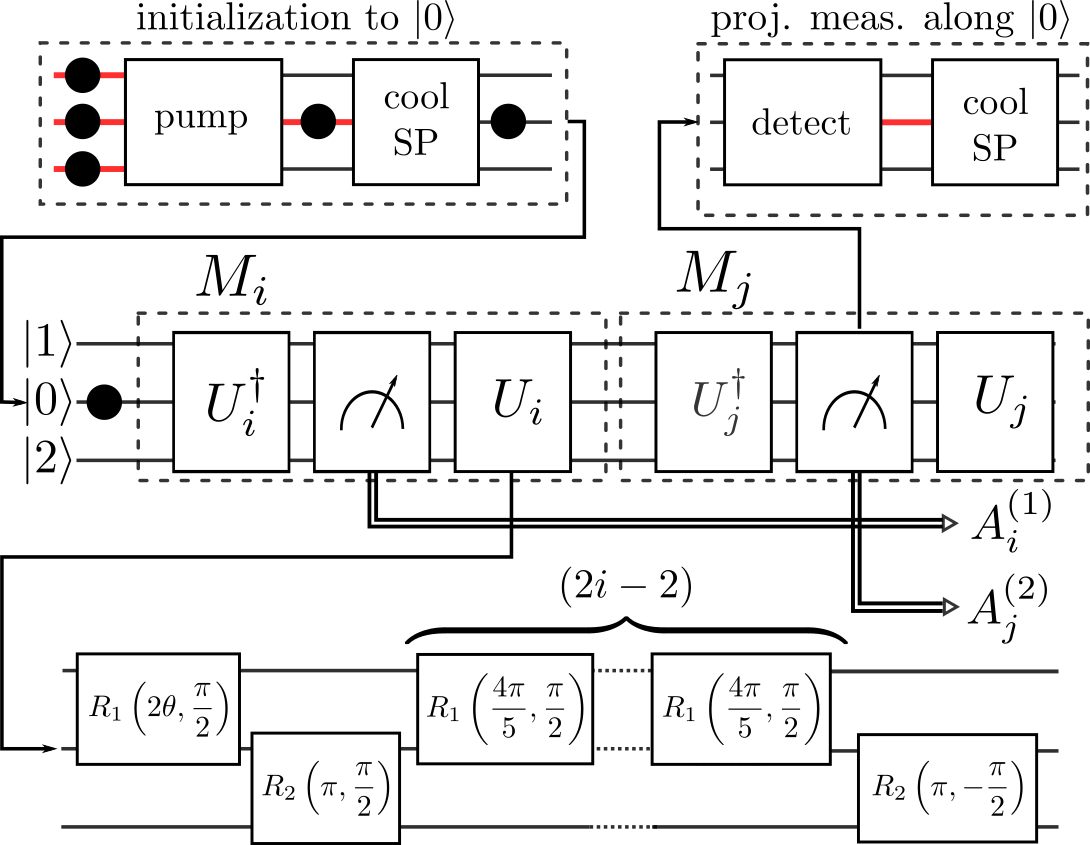}
	\caption{Sequential measurement of observables $M_i$ and $M_j$. Red lines illustrate the parts of the sequence where the ion is hot and/or outside the computational basis. In the \textit{pump} stage, the ion is Doppler-cooled and pumped into the $S_{1/2}$ manifold, which includes state $\ket{0}$. In the \textit{detect} stage, the state is projected either onto the $\{\ket{1},\ket{2}\}$ manifold (without affecting its motional state), or onto the $S_{1/2}$ manifold (heating it back to the Doppler limit). In the \textit{cool SP} stage, the $S_{1/2}$ states are ground-state cooled and pumped into $\ket{0}$, while states $\ket{1}$ and $\ket{2}$ remain unaffected. Each unitary $U_i$ is decomposed into a sequence of $(2i+1)$ coherent rotations on $\ket{0} \leftrightarrow \ket{1}$ and $\ket{0} \leftrightarrow \ket{2}$ transitions. Every sequence produces outcomes $A_i^{(1)}$ and $A_j^{(2)}$, from which we calculate the correlator $A_i^{(1)} A_j^{(2)}$. For every setting, the measurement is repeated 10,000 times.}
	\label{fig:expseq}
\end{figure}

All experiments in this work follow the sequence depicted in Fig.~\ref{fig:expseq}. We start by cooling the ion's motion close to the ground-state to suppress the effect of finite motional temperature on the fidelities of coherent control operations \cite{16Alonso}. We then optically pump the system to $\ket{0}$ and perform a measurement $M_i$ followed by a measurement of $M_{j=i\pm 1}$. Projective measurements along $\ket{0}$ are performed by applying a fluorescence state-detection pulse, followed by ground-state cooling and optical-pumping pulses \iftoggle{arXiv}{(App.~\ref{sec:cooling})}{\cite{SuppMat}}. If the ion fluoresces it is projected onto $\ket{0}$ and cooled back close to the ground-state; if it does not, it is projected onto the subspace spanned by states $\ket{1}$ and $\ket{2}$, without affecting their relative amplitudes or the motional state \cite{17Leupold}.
A projective measurement along $\ket{\psi_i} = U_i\ket{0}$ is composed of a coherent rotation $U_i^\dagger$, followed by a projective measurement along $\ket{0}$ and a rotation back $U_i$ \iftoggle{arXiv}{(App.~\ref{sec:detection}, \cite{17Leupold})}{\cite{SuppMat,17Leupold}}. This allows us to treat each measurement as a block that is executed in the same way regardless of preceding or following measurements \cite{16Cabello2}. Qutrit rotations are decomposed into individual laser pulses using Eq.~(\ref{eq:psis}), with
\begin{subequations}
	\begin{align}
	R_y (\theta) &= R_1\left(2 \theta, \tfrac{\pi}{2}\right), \\
	R_z^{2i-2}\left( \tfrac{2\pi}{5}\right) &= R_2\left(\pi,\tfrac{\pi}{2}\right) R_1^{2i-2}\left(\tfrac{4 \pi}{5},\tfrac{\pi}{2}\right) R_2\left(\pi,-\tfrac{\pi}{2}\right),
	\end{align}
\end{subequations}
where $R_{k}^n(\theta,\phi)$ ($k=1,2$) is a matrix representing the effect of a resonant pulse on the $\ket{0} \leftrightarrow \ket{k}$ transition with angle $\theta$ and phase $\phi$, repeated $n$ times \iftoggle{arXiv}{(App.~\ref{sec:transitions})}{\cite{SuppMat}}. Experimentally, these parameters are adjusted by changing the laser-pulse amplitude, duration and phase with an acousto-optic modulator.


\begin{figure}
	\includegraphics[width= 1.00 \columnwidth]{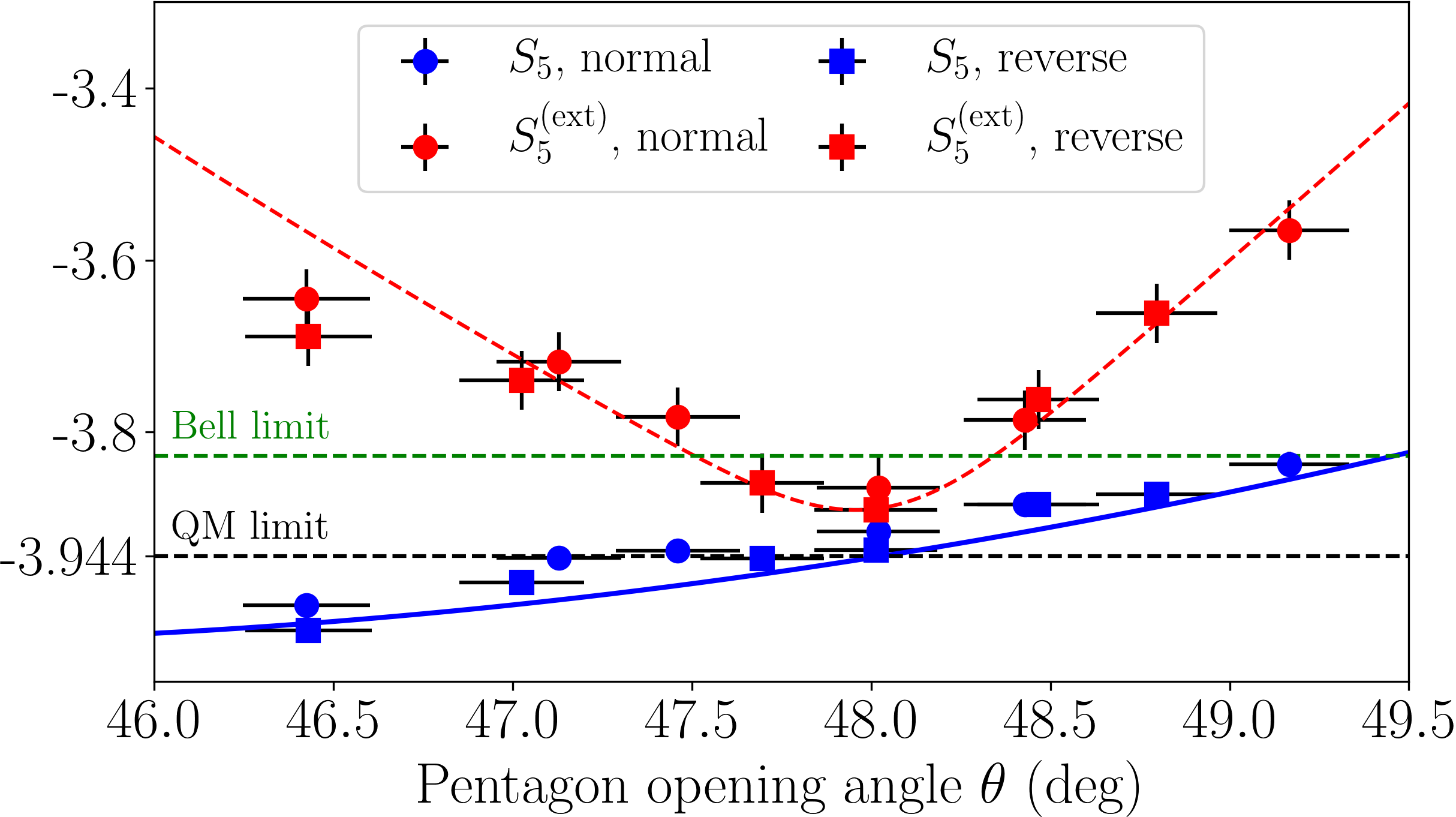}
	\caption{Results of the KCBS measurement as a function of the opening angle $\theta$. Each data point results from $10,000$ measurements on each of the five correlators $\av{A_i A_j}$, either in normal ($j=i+1$) or reverse order ($j=i-1$). Blue and dashed red lines represent theoretical expectations for $S_5$ and $S_5^\text{(ext)}$ respectively (see Fig.~\ref{fig:pentagon}, right). Note that all measurements of $S_5^\text{(ext)}$ violate the NC bound of $\bar{S}_5^\text{NC}=-3$. Error bars here and in the remainder of the paper show the standard error in the mean, with sample standard deviation obtained directly from the data \iftoggle{arXiv}{(App.~\ref{sec:dataAnalysis})}{\cite{SuppMat}}.}
	\label{fig:KCBS}
\end{figure}

\begin{table}[] 
	\centering
	\caption{Experimental results for the KCBS experiment for the points closest to the compatibility angle $\theta = \theta_5$ in both normal and reverse order in Fig.~\ref{fig:KCBS}.}
	\label{tab:kcbsResults}
	\newcolumntype{A}{ >{\centering\arraybackslash} m{0.5cm} }
	\newcolumntype{B}{ >{\centering\arraybackslash} m{1.4 cm} }
	\newcolumntype{C}{ >{\centering\arraybackslash} m{1.6cm} }
	\newcolumntype{D}{ >{\centering\arraybackslash} m{4cm} }
	\begin{tabular}{|c|l|A|A|C|C|C|}
		\hline
		\multicolumn{2}{|c|}{Order}                                                                                 & $i$          & $j$         & $\av{A_i}$           & $\av{A_j}$           & $\av{A_i A_j}$           \\ \hline
		\multicolumn{4}{|c|}{Ideal}                                                                                                & $\approx$-0.105    & $\approx$-0.105    & $\approx$-0.788                   \\ \hline
		\multicolumn{4}{|c|}{Ideal total} & \multicolumn{3}{l|}{$S_5 \approx -3.944, S_5^\text{(ext)} \approx -3.888$} \\ \hline
		\multicolumn{2}{|c|}{\multirow{6}{*}{\begin{tabular}[c]{@{}c@{}} Normal\end{tabular}}} & 1            & 2           & -0.106(10)           & -0.107(10)           & -0.786(6)                \\
		\multicolumn{2}{|c|}{}                                                                                         & 2            & 3           & -0.111(10)           & -0.092(10)           & -0.793(6)                \\
		\multicolumn{2}{|c|}{}                                                                                         & 3            & 4           & -0.107(10)           & -0.112(10)           & -0.775(6)                \\
		\multicolumn{2}{|c|}{}                                                                                         & 4            & 5           & -0.102(10)           & -0.107(10)           & -0.787(6)                \\
		\multicolumn{2}{|c|}{}                                                                                         & 5            & 1           & -0.100(10)           & -0.121(10)           & -0.774(6)                \\ \cline{3-7}
		\multicolumn{2}{|c|}{}                                                                                         & \multicolumn{2}{c|}{Total} & \multicolumn{3}{l|}{$S_5 = -3.915(14), S_5^\text{(ext)} = -3.864(34)$} \\ 
		\hline
		
		\multicolumn{2}{|c|}{\multirow{6}{*}{\begin{tabular}[c]{@{}c@{}} Reverse\end{tabular}}} & 1            & 2           & -0.113(10)           & -0.096(10)           & -0.786(6)                \\
		\multicolumn{2}{|c|}{}                                                                                         & 2            & 3           & -0.111(10)           & -0.101(10)           & -0.787(6)                \\
		\multicolumn{2}{|c|}{}                                                                                         & 3            & 4           & -0.106(10)           & -0.103(10)           & -0.784(6)                \\
		\multicolumn{2}{|c|}{}                                                                                         & 4            & 5           & -0.093(10)           & -0.118(10)           & -0.783(6)                \\
		\multicolumn{2}{|c|}{}                                                                                         & 5            & 1           & -0.102(10)           & -0.097(10)           & -0.798(6)                \\ \cline{3-7} 
		\multicolumn{2}{|c|}{}                                                                                         & \multicolumn{2}{c|}{Total} & \multicolumn{3}{l|}{$S_5 = -3.937(14), S_5^\text{(ext)} = -3.890(34)$} \\ \hline
		
	\end{tabular}
\end{table}

In the KCBS study we scan the degree of incompatibility between observables by changing the pentagon opening angle $\theta$ and measuring each pair of observables 10,000 times in both normal and reverse order. We determine the experimental value of $\theta$ using
\begin{equation}
\theta = \frac{1}{2}\arccos(\sum_{i=1}^{N}\frac{\av{A_i^{(1)}}}{N}).
\end{equation}
The measured witnesses $S_5(\theta)$ and $S_5^\text{(ext)}(\theta)$ are displayed in Fig.~\ref{fig:KCBS}, together with theoretical expectations for an ideal experiment. Table \ref{tab:kcbsResults} shows the results of this procedure for the value of $\theta$ measured to be closest to $\theta_5\approx 48^{\circ}$ in each scan. The results of $S_5(\theta_5)$ exhibit a systematic shift of $1.6$ standard deviations from the ideal QM prediction. This can be attributed to imperfections in qutrit rotations, primarily due to vibrations of the closed-cycle cryostat where the ion trap sits. For the extended witness $S_5^\text{(ext)}(\theta_5)$, statistical errors dominate. The data point closest to compatibility violates the KCBS Ineq.~(\ref{eq:KCBS}) by 65 (67) standard deviations in the normal (reverse) order. In addition, all measured data points violate the extended KCBS Ineq.~(\ref{eq:extKCBS}) by up to 25 standard deviations \footnote{The complete raw dataset is publicly available from an open repository on \url{http://www.tiqi.ethz.ch/publications-and-awards/public-datasets.html}.}. 

Recent theoretical developments allow for a unified treatment of different NC inequalities and consequently justify a direct comparison between contextuality and Bell tests \cite{13Cabello}. Within the formalism of exclusivity structures \iftoggle{arXiv}{(App.~\ref{sec:exclusivity})}{\cite{SuppMat}}, KCBS and its extensions (odd $N$-cycle NC inequalities \cite{13Araujo}) correspond to the most fundamental exclusivity scenarios, which are building blocks of all other NC inequalities. Bell experiments (even $N$-cycle NC inequalities) produce correlations that cannot saturate those available due to their exclusivity graph. The largest amount of contextuality available to Bell scenarios would correspond to $S_5\approx -3.828$ (green dashed line in Fig.~\ref{fig:KCBS}, \iftoggle{arXiv}{App.~\ref{sec:exclusivity}}{\cite{SuppMat}}). Close to compatibility we can resolve values of $S_5$ surpassing this bound

In a second set of experiments we expand the above procedure to correlation measurements between any odd number $N>5$ of states on the ``qutrit sphere''. Generalizing the KCBS construction given in Eq.~(\ref{eq:psis}), we define the $N$-gon states by
\begin{align}	\label{eq:psisN}
\ket{\psi_i} = U_i \ket{0}	\qq{with}
U_i= R_z^{(i-1)(N-1)/2}( \tfrac{2\pi}{N}) R_y(\theta).
\end{align}
Pairwise compatibility $\bra{\psi_{i}}\ket{\psi_{i\pm1}} = 0$ occurs when $\theta = \theta_N = \arccos{\sqrt{\frac{\cos(\frac{\pi}{N})}{1+\cos(\frac{\pi}{N})}}}$. The $N$-cycle NC inequality \cite{10Cabello, 11Liang} reads:
\begin{equation} 
\label{eq:KCBSN}
S_N=\sum_{i=1}^{N}\av{A_{i}A_{i+1}}\geq \bar{S}_N^\text{NC}=-N+2.
\end{equation}
Measurements on the initial state $\ket{0}$ violate this inequality maximally, leading to
\begin{equation} 
\label{eq:KCBSQMN}
S_N \geq \bar{S}_N^\text{QM}=\frac{N - 3N\cos(\pi/N)}{1+\cos(\pi/N)}.
\end{equation}
Finally, an inequality for the extended witness $S_N^\text{(ext)}$ can be derived in full analogy to Ineq.~(\ref{eq:extKCBS}) \iftoggle{arXiv}{(App.~\ref{sec:ngonDetails}, \cite{15Kujala})}{\cite{SuppMat}}.

In order to make a connection with chained Bell tests \cite{11Abramsky,13Sadiq}, we use the ``Contextual Fraction'' (CF),
\begin{equation}\label{eq:CF}
\text{CF}_N = \frac{S_N - \bar{S}_N^\text{NC}}{\bar{S}_N^\text{NS}-\bar{S}_N^\text{NC}},
\end{equation}
to quantify the strength of non-classical correlations. Here, $\bar{S}_N^\text{NS}$ is the minimum value of $S_N$ allowed for non-signaling measurements. For $N$-cycle NC inequalities, $\bar{S}_N^\text{NS}=-N$, which is also the algebraic limit of the expression. When positive, the value of CF quantifies the potential performance of measurement-based quantum computers \cite{17Abramsky} and can be used to constrain possible extensions of QM \cite{12Amselem,*11Colbeck}. A property of $N$-cycle NC inequalities is that  $\text{CF}_N \rightarrow 1$ as $N\rightarrow \infty$. In other words, the system tends to become fully contextual as the number of observables increases. Chained Bell experiments have observed contextuality with $N$ up to 90, measuring CF as large as $\text{CF}_{36} = 0.874(1)$ \cite{15Christensen}. Here we complement those studies by measuring odd $N$-cycle NC inequalities up to $N=121$, while quantifying systematic effects which can compromise experiments on photonic systems \cite{18Smania}.


The number of pulses and duration for these experiments both grow as $N^2$. In order to shorten the experimental run time we concatenate $U_j^{\dagger} U_i$ to $U_{i-j}$, rather than performing the full pulse sequence corresponding to $U_j^{\dagger} U_i$ \iftoggle{arXiv}{(App.~\ref{sec:ngonDetails})}{\cite{SuppMat}}. This precludes the block-like structure of individual measurements \cite{16Cabello2}, but leads to relevant time and infidelity reductions for large $N$. All measurements were performed in normal (as opposed to reverse) order, with every correlator measured $10,000$ times.

\begin{figure}[ht]
	\includegraphics[width= 1.00 \columnwidth]{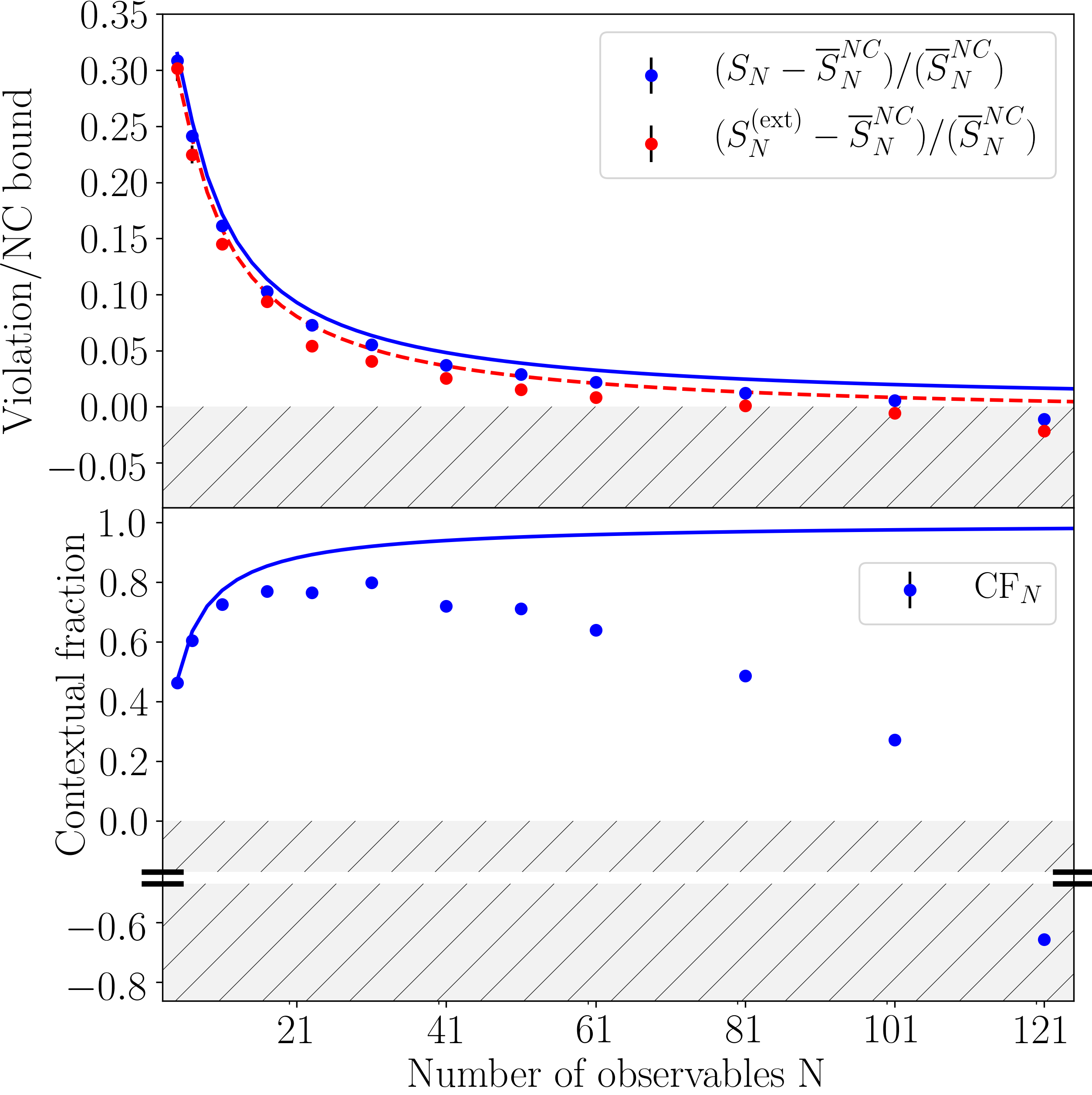}
	\caption{Results of measurements of $N$-cycle witnesses. Solid and dashed lines show QM expectations for relevant quantities (red dashed line includes shot noise), and hashed regions below 0 correspond to results explainable by classical models. The top plot illustrates the fractional gap between quantum and classical witnesses, which decreases for large $N$. Our data shows contextuality up to $N=101$ ($N=61)$ for $S_N$ ($S_N^{\text{(ext)}}$). The bottom plot shows the calculated contextual fraction. Ideally the system becomes fully contextual at $N\rightarrow\infty$, but experimental imperfections lead to $\text{CF}<0$ for large enough $N$. We measure $\text{CF}_{31} = 0.800(4)$. Error bars are generally smaller than the point size.}
	\label{fig:fraction}
\end{figure}

The measurement results are shown in Fig.~\ref{fig:fraction}. We identify stronger-than-classical correlations in all prepared scenarios up to $N=101$ ($N=61$) for the bare (extended) witness. Beyond that, our results are consistent with NC models. The largest measured value is $\text{CF}_{31}=0.800(4)$. We have not found a consistent theory model for calculating the CF in the presence of finite signaling, so the plot includes only the results for the bare witnesses. The complete table of results, as well as further experimental details, are available in \iftoggle{arXiv}{App.~\ref{sec:ngonDetails}}{\cite{SuppMat}}. To our best knowledge, these results show contextuality in a system with the largest number of observables (101) of any experiment reported up to this date. Moreover, the measured contextual fraction is larger than for any other experiment closing the detection loophole \cite{17Tan}.

The experimental difficulty of resolving quantum from classical correlations for a large number of observables is illustrated in the bottom plot of Fig.~\ref{fig:fraction}. A decrease in the contextual fraction corresponds to a loss in the visibility of quantum correlations. While QM predicts that $\text{CF}_N$ will approach unity as $N$ increases, any finite error rate per measurement will cause $\text{CF}_N<0$ for sufficiently large $N$. This can be interpreted as classical behavior emerging from a quantum system \cite{07Kofler,14Jeong}: whereas the strength of correlations between observables for $N \leq 101$ cannot have a classical origin, the correlations we measure for 121 observables could have been produced by a system of 121 classical coins. This transition towards classicality does not originate from increased interactions with an environment (measured decoherence in our system occurs over timescales much longer than the duration of our experimental sequences, see \iftoggle{arXiv}{App.~\ref{sec:coherence}}{\cite{SuppMat}}) and is not due to an increase in the macroscopicity of the system (the Hilbert-space dimension is always 3). Instead, we find classical expectations when we do not have sufficient experimental control to resolve measurement directions from one another. This possibility was first discussed in Ref.~\cite{07Kofler}, where it is proven that measurement imperfections in an otherwise perfectly quantum system can give rise to classical behavior. It was later shown that measurement errors do not lead to a quantum-to-classical transition if the final projection is coarsened, whereas they do when measurement references are coarsened \cite{14Jeong}, i.e. when unitary operations are non-ideal, which is consistent with our observations. Although this effect is also present in previous chained Bell tests \cite{15Christensen,17Tan}, the link to quantum-to-classical transitions has not been discussed in this context before.


The experiments we have performed suggest that non-classical correlations as strong as predicted by quantum mechanics can be observed in nature. Using this as a starting point, experiments should aim at closing the compatibility loophole to disprove all possible NC models. This can be accomplished by using multiple entangled qutrits \cite{11Cabello}. A further interesting study would be to carry out experiments using an operational definition of contextuality which addresses the compatibility loophole and does not require sharp measurements \cite{05Spekkens,*17Kunjwal}. Initial work has already been carried out with photonic systems to study the monogamy relation between contextuality and non-locality \cite{16Zhan}, as well as alternative definitions of contextuality \cite{16Mazurek}. Comparable experiments on a trapped-ion platform would facilitate closing the detection loophole.


\begin{acknowledgments}
We thank Renato Renner, Ravi Kunjwal, Myungshik Kim, and Hyunseok Jeong for discussions, and Chiara Decaroli for comments to the manuscript. We acknowledge support from the Swiss National Science Foundation under grant no. 200021 134776. The research is partly based upon work supported by the Office of the Director of National Intelligence (ODNI), Intelligence Advanced Research Projects Activity (IARPA), via the U.S. Army Research Office grant W911NF-16-1-0070. The views and conclusions contained herein are those of the authors and should not be interpreted as necessarily representing the official policies or endorsements, either expressed or implied, of the ODNI, IARPA, or the U.S. Government. The U.S. Government is authorized to reproduce and distribute reprints for Governmental purposes notwithstanding any copyright annotation thereon. Any opinions, findings, and conclusions or recommendations expressed in this material are those of the author(s) and do not necessarily reflect the view of the U.S. Army Research Office. AC acknowledges support from Project No.\ FIS2014-60843-P, ``Advanced Quantum Information'' (MINECO, Spain), with FEDER funds, the FQXi Large Grant ``The Observer Observed: A Bayesian Route to the Reconstruction of Quantum Theory,'' and the project ``Photonic Quantum Information'' (Knut and Alice Wallenberg Foundation, Sweden).
\end{acknowledgments}

\textbf{Author Contributions:} Experimental data were taken by CZ, MM and JA, using an apparatus primarily built up by FML and JA, and with significant contributions from MM and CZ. Data analysis was performed by MM, JA, and AC. The paper was written by JA, MM, JPH and AC, with input from all authors. Experiments conceived by AC, MM, JA, and JPH.

The authors declare that they have no competing financial interests.


\iftoggle{arXiv}{
	\clearpage
	{\LARGE\bfseries Supplementary Material}
}{}

\appendix

\section{Qutrit transitions}
\label{sec:transitions}
Qutrit states are encoded into Zeeman sub-levels of a $^{40}$Ca$^+$ ion as follows:
\begin{align*}
	\ket{0} =& \ket{S_{1/2},\  m_J = -1/2}, \\
	\ket{1} =& \ket{D_{5/2},\  m_J = -3/2}, \\
	\ket{2} =& \ket{D_{5/2},\  m_J = -1/2}.
\end{align*}
An external magnetic field of $|B| \approx\SI{3.73}{G}$ splits the $\ket{0} \leftrightarrow \ket {1}$ and $\ket{0} \leftrightarrow \ket {2}$ transitions by $\approx\SI{6.27}{MHz}$. These transitions are driven by linearly polarized laser pulses at $\lambda \approx\SI{729}{nm}$ propagating at an angle of $45^{\circ}$ to the quantization axis and adjusted to be resonant with the transition of choice. We operate at low laser intensities to keep AC Stark shifts below \SI{100}{Hz}. This allows us to, up to a good approximation, treat the laser pulses as inducing single-qubit rotations in a qutrit space:
\begin{subequations}	\label{eq:RotationOperators}
	\begin{align}
		R^{(1)}(\theta,\phi)&=\left(\begin{array}{ccc}
			\cos(\frac{\theta}{2}) & -\I \Exp{-\I\phi} \sin(\frac{\theta}{2}) & 0 \\[0.5ex]
			-\I \Exp{\I\phi} \sin(\frac{\theta}{2}) & \cos(\frac{\theta}{2}) & 0 \\[0.5ex]
			0 & 0 & 1
		\end{array}\right),	\\
		R^{(2)}(\theta,\phi)&=\left(\begin{array}{ccc}
			\cos(\frac{\theta}{2}) & 0 & -\I \Exp{-\I\phi} \sin(\frac{\theta}{2}) \\[0.5ex]
			0 & 1 & 0 \\[0.5ex]
			-\I \Exp{\I\phi} \sin(\frac{\theta}{2}) & 0 & \cos(\frac{\theta}{2})
		\end{array}\right).
	\end{align}
\end{subequations}

\section{Qutrit coherence times}
\label{sec:coherence}
Using Ramsey techniques, we determine coherence times of $\sigma_t \approx \SI{1.6}{ms}$ for the $\ket{0} \leftrightarrow \ket {1}$ and $\ket{0} \leftrightarrow \ket {2}$ transitions, and $\sigma_t \approx \SI{7}{ms}$ for the $\ket{1} \leftrightarrow \ket {2}$ transition. Here we assumed for simplicity that the noise is Gaussian and slow compared to the timescale of a single experiment, causing a Ramsey experiment with wait time $\tau$ to lose contrast as $C = \Exp{-\tau^2/(2 \sigma_t^2)}$ \cite{98Wineland2}. This reveals a noise component of $\approx\SI{230}{Hz}$ (Full-Width Half-Maximum, FWHM) common to both the $\ket{0} \leftrightarrow \ket {1}$ and $\ket{0} \leftrightarrow \ket {2}$ transitions, which we believe originates due to Doppler shifts transmitted onto the ion from vibrations of the closed-cycle cryocooler. The differential noise has a width of $\approx\SI{50}{Hz}$ (FWHM), which we associate with $B$-field fluctuations of <\SI{5}{\micro G} (FWHM) and slow drifts. The latter result in changes of transition frequencies ($\sim\SI{100}{Hz}$ on time scales of minutes), which we automatically re-calibrate every \SI{30}{s} with \SI{10}{Hz} resolution.\\

\section{Ion cooling}
\label{sec:cooling}
At the start of every experiment the ion is optically pumped to $\ket{0}$ and cooled close to the motional ground state. The cooling is done in two steps. First, a Doppler-cooling sequence brings all three oscillation modes (one parallel, and two perpendicular to the trap axis) to a thermal state with mean phonon occupation $n_{\text{th}} \approx 5$. Then we further cool the mode of motion parallel to the trap axis to $n_{\text{th}} \approx 0.2$ using electromagnetically-induced transparency (EIT) cooling \cite{16Alonso}. Qutrit transitions are driven with a laser beam propagating along the trap axis, deeming the influence of radial motion negligible.

The second instance of cooling occurs during state detection. Standard on-resonance fluorescence detection techniques cause motional heating due to photon recoil in case of a bright detection. To circumvent this problem, detection is conducted off-resonantly with settings close to those of Doppler cooling \cite{17Leupold}. While this decreases the observed signal and hence lengthens detection pulses, it also brings the temperature of the bright ion close to the Doppler limit. Afterwards, we repeat the EIT cooling sequence described above to get back to $n_{\text{th}} \approx 0.2$. In case of a dark detection, the state of the ion is unaffected by the detection or EIT beams. This leads to negligible dependence of the ion's motional state on the first measurement, since a dark ion is subject to motional heating at a rate of $\sim \SI{200}{quanta/s}$.

\section{Qutrit detection}
\label{sec:detection}
Fluorescence detection is implemented by shining a laser beam at \SI{397}{nm}, which connects the $S_{1/2}$ states with a short-lived $P_{1/2}$ level, together with a repumping beam at \SI{866}{nm} (Fig.~\iftoggle{arXiv}{\ref{fig:expseq}}{\ref{m-fig:expseq}}, \cite{17Leupold}). If the ion fluoresces (``bright detection``) it is projected onto the $S_{1/2}$ manifold, while no fluorescence (``dark detection``) projects it onto the $D_{5/2}$ manifold.

During a dark detection, the lack of physical interaction between ion and laser ensures that the initial qutrit state $\rho_{\text{in}}$ is related to the post-measurement state $\rho_{\text{out}}$ via
\begin{equation}
\rho_{\text{out}} = \frac{P_\text{D} \rho_{\text{in}} P_\text{D}}{\text{tr}(P_\text{D} \rho_{\text{in}})},
\end{equation}
where $P_\text{D} = (\ket{1}\bra{1} + \ket{2}\bra{2})$ is the projection operator onto the dark states. In other words, a dark detection implements a L\"uders projective measurement \cite{50Luders}. A bright detection, on the other hand, projects the ion onto a mixture of $\{\ket{S_{1/2}, m_J = -1/2},\ket{S_{1/2}, m_J = +1/2}\}$ states. We therefore complete the detection sequence with a spin-polarizing (SP) $\sigma$-polarized pulse at \SI{397}{nm}, which pumps all the $S_{1/2}$ population into $\ket{0} = \ket{S_{1/2},\  m_J = -1/2}$. This makes a bright measurement a L\"uders measurement with projection operator $P_\text{B} = \ket{0}\bra{0}$.

Bright states are discriminated from dark states by thresholding the number of photon counts collected within a detection window. In a typical experiment, we collect an average of $\approx 25$ photons from a bright ion, with an average background of $\approx 1$ photon, during a window of $\approx\SI{200}{\micro s}$. The optimal threshold is set close to the crossing point between the histograms resulting from bright and dark states \cite{17Leupold}. We estimate a bright (dark) detection error of $\approx\num{2e-5}$ ($\approx\num{1e-4}$). Dark-detection errors are dominated by spontaneous decay of $D_{5/2}$ states into the $S_{1/2}$ ground state ($\tau_{\text{decay}} \approx \SI{1.2}{s}$).

\section{Data collection and analysis}
\label{sec:dataAnalysis}
Consider an experiment with $N$ observables (for KCBS, $N=5$). For each opening angle setting $\theta_{\text{set}}$ we perform a correlation measurement (Fig.~\iftoggle{arXiv}{\ref{fig:expseq}}{\ref{m-fig:expseq}}) along a pair of directions $(\ket{\psi_i},\ket{\psi_{j}})$, with results $(A_i^{(1)} = \pm 1$, $A_j^{(2)} = \pm 1)$. Results +1 correspond to a bright ion and results -1 correspond to a dark ion, and the superscript in brackets denotes whether the observable is measured first or second. We also calculate the correlation $A_i^{(1)}A_j^{(2)}$. Each experiment is repeated 10,000 times, and the average results ($\av{A_i^{(1)}}, \av{A_j^{(2)}}, \av{A_i^{(1)}A_j^{(2)}}$) are extracted. We collect data for $N$ observable pairs in one of two possible orders:
\begin{align}
	\text{normal order:}& \quad (i,j) = (i,i+1),\quad i=1,\ldots,N, \\
	\text{reverse order:}& \quad (i,j) = (i,i-1),\quad i=1,\ldots,N.
\end{align}

We then evaluate the witness $S_N$ starting from Eq.~(\iftoggle{arXiv}{\ref{eq:KCBSN}}{\ref{m-eq:KCBSN}}):
\begin{align}
	\text{normal order:}& \quad S_N = \sum_{i=1}^N \av{A_i^{(1)} A_{i+1}^{(2)}}, \\
	\text{reverse order:}& \quad S_N = \sum_{i=1}^N \av{A_{i+1}^{(1)} A_{i}^{(2)}}.
\end{align}
We also estimate the pentagon opening angle $\theta$ using
\begin{equation}
\theta = \frac{1}{2}\arccos(\sum_{i=1}^{N}\frac{\av{A_i^{(1)}}}{N}).
\end{equation}
The incompatibility term $\epsilon$ is evaluated according to
\begin{equation}
\epsilon = \sum_{i=1}^N \epsilon_i = \sum_{i=1}^N |\av{A_i^{(1)}} - \av{A_i^{(2)}}|,
\end{equation}
and the extended witness is given by $S_N^{\text{(ext)}} = S_N + \epsilon $ (Eq.~(\iftoggle{arXiv}{\ref{eq:extKCBS}}{\ref{m-eq:extKCBS}})).
For each average ($\av{A_i^{(1)}}, \av{A_j^{(2)}}, \av{A_i^{(1)}A_j^{(2)}}$) we evaluate the sample standard deviation and use it to compute the standard error in the mean. We then propagate the standard errors of $S_N, \theta, \epsilon$ and $S_N^{\text{(ext)}}$ assuming independent errors. These standard errors are plotted as error bars in Figs.~\iftoggle{arXiv}{\ref{fig:KCBS}}{\ref{m-fig:KCBS}} and \iftoggle{arXiv}{\ref{fig:fraction}}{\ref{m-fig:fraction}}. We note that $\epsilon$ is defined as necessarily positive and its distribution is non-gaussian, hence the standard error of $\epsilon$ cannot always be treated as a confidence interval.

\section{Measured observables}
\label{sec:observables}
Consider operators $P_\text{B} = \ket{0}\bra{0}$ and $P_\text{D} = \ket{1}\bra{1} + \ket{2}\bra{2}$, which project the ion onto a bright and dark state respectively. We define the observable $M_i$ of measurement along $\ket{\psi_i} = U_i \ket{0}$ as $M_i = P_{\text{B},i} - P_{\text{D},i}$, where
\begin{align}
	P_{\text{B},i} &= U_i P_\text{B} U_i^{\dagger} = \ket{\psi_i}\bra{\psi_i},\\
	P_{\text{D},i} &= U_i P_\text{D} U_i^{\dagger}.
\end{align}
The outcome of measurement $M_i$ is denoted as $A_i = \pm1$. With these definitions, the outcome $A_i=+1$ corresponds to projection onto $P_{\text{B},i}$, while $A_i=-1$ corresponds to projection onto $P_{\text{D},i}$. Note that $[M_i,M_{i+1}]=0$ when $\bra{\psi_i}\ket{\psi_{i\pm1}} = 0$ and that in QM two observables are compatible when their operators commute. If we consider states $\ket{\psi_i}$ on a real ``qutrit sphere'' (Fig.~\iftoggle{arXiv}{\ref{fig:pentagon}}{\ref{m-fig:pentagon}}), this condition is equivalent to orthogonality between vectors.

\section{Theoretical predictions for KCBS witnesses}
\label{sec:theory}
Consider a sequence of measurements $M_i$, $M_{i\pm 1}$ as defined in App.~\ref{sec:observables}, with $U_i$ defined as in Eq.~(\iftoggle{arXiv}{\ref{eq:psis}}{\ref{m-eq:psis}}) and the initial state $\rho_{\text{in}} = \ket{0}\bra{0}$. Their correlation can be evaluated to:
\begin{eqnarray}
\notag \av{A_i^{(1)} A_{\pm1}^{(2)}} &=& \text{tr} \big( M_i M_{i\pm1} \rho_{\text{in}})\\
&=& \dfrac{1}{8} (3-\sqrt{5} + (5+\sqrt{5})\cos(4\theta)\big).
\end{eqnarray}
Then the KCBS witness can be simply calculated as $S_5(\theta) = 5\av{A_i^{(1)} A_{i\pm1}^{(2)}}$ (Eq.~(\iftoggle{arXiv}{\ref{eq:KCBS}}{\ref{m-eq:KCBS}})). This is plotted in solid blue in Fig.~\iftoggle{arXiv}{\ref{fig:pentagon}}{\ref{m-fig:pentagon}}. Note in particular that the minimum value of $S_5(\theta)$ is obtained when $\theta = \frac{\pi}{2}$, where $S_5 = \frac{5}{4}(-\sqrt{5}-1) \approx -4.045$. At the point of compatibility, where $[M_i,M_{i+1}]=0$, which corresponds to $\theta=\theta_5 = \arccos (5^{-1/4})$, we obtain $S_5 = 5 - 4 \sqrt{5} \approx -3.944$.

In order to evaluate the extended KCBS witness $S_5^{\text{(ext)}}(\theta)$ we first evaluate the expectation value of $A_i$ as
\begin{equation}
\av{A_i^{(1)}} = \text{tr} ( M_i \rho_{\text{in}}) = \cos (2\theta),
\end{equation}
while the post-measurement state is given by
\begin{equation}
\rho_{i} = P_{\text{B},i}\rho_{\text{in}} P_{\text{B},i} + P_{\text{D},i}\rho_{\text{in}} P_{\text{D},i}.
\end{equation}
The average outcome of measurement $M_{i\pm1}$ is then given by
\begin{equation}
\av{A_{i\pm1}^{(2)}} = \text{tr} ( M_{i\pm1} \rho_{i}).
\end{equation}
Due to the symmetry of the problem, $\av{A_{i+1}^{(2)}}$ is expected to be the same for all $i$, and hence$\av{A_{i+1}^{(2)}} = \av{A_{i}^{(2)}}$. We use this to evaluate the extension term:
\begin{eqnarray}
\notag \epsilon_i &=& \abs{\av{A_{i}^{(1)}}-\av{A_{i}^{(2)}}} = \abs{\av{A_{i}^{(1)}}-\av{A_{i+1}^{(2)}}}\\ &=&\dfrac{1}{16} \abs{(5 - \sqrt{5} + 5 (3 + \sqrt{5}) \cos{(2\theta)}) \sin^2{(2\theta)}}.
\end{eqnarray}
The extended KCBS witness $S_5^{\text{(ext)}} = S_5 + \sum_{i=1}^5 \epsilon_i$ is plotted in Fig.~\iftoggle{arXiv}{\ref{fig:pentagon}}{\ref{m-fig:pentagon}} in solid red. Note that at the point of compatibility $\epsilon_i=0$ and $S_5^{\text{(ext)}} = S_5 = 5 - 4 \sqrt{5}$, while for all other values of $\theta$ we have $S_5^{\text{(ext)}}>\bar{S}_5^\text{QM}$. 

Aside from penalizing systematic effects, $\epsilon_i$ also accounts for finite sample size. This is because the sample mean of $\epsilon_i$ is a biased estimator of the population mean of $\epsilon_i$. Consider an experimental run with $n$ measurement repetitions. The measured outcomes follow a normal distribution $Y_{\av{A_i}} \propto \mathcal{N}(\av{A_i},\sigma^2_{A_i})$, where $\mathcal{N}(\mu,\sigma)$ denotes a normal distribution with mean $\mu$ and variance $\sigma^2$. Then $Y_{\av{A_i}-\av{A_{i+1}}} \propto \mathcal{N}(\av{A_i}-\av{A_{i+1}},\sigma^2_{A_i}+\sigma^2_{A_{i+1}})$. Finally, $Y_{\epsilon_i} = \mathcal{F} (\av{A_i}-\av{A_{i+1}},\sigma^2_{A_i}+\sigma^2_{A_{i+1}})$, where $\mathcal{F}(\mu,\sigma^2)$ is the so-called ``folded normal distribution``, obtained by taking the absolute value of a gaussian with mean $\mu$ and variance $\sigma^2$. The mean of a folded normal distribution is \cite{61Leone}
\begin{equation}
\mu_\text{F}\left(\mu,\sigma\right) = \sigma \sqrt{\dfrac{2}{\pi}} \text{exp}\left(\dfrac{-\mu^2}{2\sigma^2}\right) - \mu \  \text{erf}\left(\dfrac{-\mu}{\sqrt{2\sigma^2}}\right),
\end{equation}
where $\text{erf}$ is the error function. This mean represents an expectation value $\text{E}(\epsilon_i)$ of $\epsilon_i$ in an experiment with $n$ repetitions:
\begin{eqnarray}
\text{E}\left(\epsilon_i\right) &=& \ \mu_\text{F}\left(\epsilon_i, \sigma^2_{A_i}+\sigma^2_{A_{i+1}}\right),\\
\sigma^2_{A_i} &=& \ \dfrac{1-\av{A_i}}{n}.
\end{eqnarray}
The expression for the variance $\sigma^2_{A_i}$ represents the shot-noise contribution, hence we refer to this expectation value as ``theoretical prediction with shot noise''. $\text{E}(S_5^{\text{(ext)}}) = S_5 + \sum_{i=1}^5 \text{E}(\epsilon_i)$ for $n=10,000$ is plotted in Fig.~\iftoggle{arXiv}{\ref{fig:pentagon}}{\ref{m-fig:pentagon}} in dashed red. We see that the effect of shot noise is significant when $\epsilon_i \approx 0$, where it predicts a gap of size $\sqrt{2(1-\cos{(2\theta_5)})/(\pi n)}$ between the sample mean and the population mean of $S_5^{\text{(ext)}}$.

\section{Relevance of KCBS and odd cycle NC inequalities}
\label{sec:exclusivity}

There are three different perspectives from which the KCBS and the $N$ odd-cycle non-contextuality inequalities (with $N\ge5$; the case $=5$ is the KCBS inequality) are of fundamental importance for understanding the power of quantum systems. Each of these perspectives corresponds to a different approach for investigating the quantum vs classical advantage.

The first perspective focuses on {\em the quantum system that produces the quantum advantage}. In this respect, the KCBS and the other odd $N-$cycle NC inequalities are special because all of them achieve their maximal quantum violation using a {\em qutrit}, which is the simplest quantum system that produces contextuality \cite{66Bell,67Kochen}. This is in contrast with the case of the CHSH and the rest of even $N-$cycle NC inequalities (called chained Bell inequalities when tested on pairs of systems \cite{70Pearle,90Braunstein}), whose maximal quantum violation requires either a ququart, i.e. a four-dimensional quantum system, or a pair of qubits.
Moreover, the KCBS inequality is the simplest (i.e. the one requiring the smallest number of observables) non-contextuality inequality violated by a qutrit.

The second perspective emphasizes {\em the scenario in which the quantum advantage occurs}. A scenario is defined by a set of observables, each with a certain number of possible outcomes, having certain relations of compatibility among them. These relations are typically represented by a graph called the {\em compatibility graph}, in which vertices represent observables and edges connect compatible observables. The name $N$-cycle inequality refers to the fact that the corresponding scenario has $N$ observables whose compatibility relations are represented by a cycle of $N$ vertices. For each scenario, the correlations between compatible measurements define a set whose points are vectors of probabilities. Geometrically, the classical set is a polytope whose facets define tight inequalities that are necessary and sufficient conditions for the existence of a classical model. For certain scenarios, the quantum set is larger than the classical set. However, only a few scenarios have been exhaustively explored \cite{82Fine,03Sliwa,04Collins,13Araujo,16Lopez} in the sense that we know all the classical inequalities and all the maximal quantum violations. Among these scenarios, the $N \ge 4$ cycle non-contextuality scenario (where there are $N$ observables, each with $2$ outcomes, such that observable $j$ is compatible with observable $j+1$ with the sum modulo $N$) is the only one which is symmetric, i.e. with all tight inequalities and all maximal quantum violations being of the same type \cite{13Araujo}. These inequalities are precisely the $N \ge 4$ cycle non-contextuality inequalities. As non-contextuality inequalities, all of them are tight. However, the case $N$ even can be tested with pairs of particles and then the corresponding non-contextuality inequalities are also Bell inequalities, the so-called chained Bell inequalities \cite{70Pearle,90Braunstein}. Notice, however, that the chained Bell inequalities are {\em not} tight Bell inequalities (i.e. they do not correspond to facets of the {\em local} polytope) \cite{13Araujo}.

The third perspective focuses on {\em the graph of exclusivity responsible for the quantum advantage}. In the graph-theoretic approach to quantum theory introduced in \cite{14Cabello}, every classical inequality is first converted into an inequality in which the quantity bounded is a sum of probabilities of events (i.e. state transformations produced by an ideal measurement), and then associated to a graph in which vertices represent the events in this quantity and edges connect events that are exclusive (i.e. that correspond to different outcomes of the same ideal measurement). This graph is called the exclusivity graph of the inequality. 
The exclusivity graph of (the events of) an inequality should not be confused with the compatibility graph of (the observables) of an scenario.

The first result of the graph-theoretic approach is that an exclusivity graph admits a quantum realization whose probabilities cannot be explained classically {\em if and only if} the graph has, as induced subgraphs, odd $M$-cycles with $M\ge5$ or their complements \cite{14Cabello}. Relevant to this study is the fact that the exclusivity graph of the $M$ events needed to test the odd $N$-cycle NC inequalities is precisely an $M$-cycle with $M=N$.

The second result of the graph-theoretic approach is that, for any exclusivity graph, the maximum value allowed by QM is given by a characteristic number of the graph. The odd $N \geq 5$-cycle NC inequalities happen to saturate each of these maxima \cite{14Cabello} and hence saturate also the strength of correlations allowed by QM, which is given by Eq.~(\iftoggle{arXiv}{\ref{eq:KCBSQMN}}{\ref{m-eq:KCBSQMN}}):
\begin{equation} 
\bar{S}_M^\text{QM}=\frac{M - 3M\cos\rbra{\frac{\pi}{M}}}{1 + \cos\rbra{\frac{\pi}{M}}}.
\end{equation}

Together, these two results imply that, among all possible non-contextuality inequalities, the odd $N \geq 5$-cycle NC inequalities have fundamental importance for understanding the differences between quantum and classical resources. Moreover, the exclusivity graph with the smallest number of vertices featuring genuinely quantum probabilities is the pentagon \cite{14Cabello}, which is exactly the exclusivity graph of the five events needed to test the KCBS inequality. 

Although the chained Bell inequalities (and other Bell inequalities) contain events associated with $N$-cycle exclusivity graphs, it has been proven \cite{13Sadiq,17Rabelo} that no Bell inequality contains events allowing the maximum quantum value for any $N$-cycle exclusivity graph. It has also been proven \cite{17Rabelo} that the quantum maximum of the odd $N \geq 5$-cycle exclusivity graph for Bell scenarios is achieved for the chained Bell inequalities for $(N-1)/2$ observables per party and is given by
\begin{equation}
\bar{S}_M^\text{Bell} = M - 4\sbra{\frac{1}{2} + \frac{M-1}{4}\rbra{1 + \cos\rbra{\frac{\pi}{M-1}}}}.
\end{equation}
For example, for the pentagon, the quantum maximum within Bell scenarios is achieved in the scenario with two dichotomic observables per party, i.e. in the CHSH scenario \cite{69Clauser}. A direct comparison between the QM expectations for the CHSH and KCBS scenarios shows that correlations between observables are stronger in the latter: $\bar{S}_5^\text{Bell}\approx -3.828 > \bar{S}_5^\text{KCBS} \approx -3.944$. Although the quantum realization for the CHSH scenario takes place in a Hilbert space of dimension four while for the KCBS scenario takes place in dimension three, measurements in the former are restricted to be bipartite, which accounts for the weaker correlations.

Finally, the third important result of the graph-theoretic approach is that, for a given graph of exclusivity, the set of classical probabilities is the so-called stable set polytope of the graph, while the set of quantum probabilities is the so-called theta body of the graph \cite{14Cabello}. Among all the basic graphs needed for genuinely quantum correlations, the pentagon is the one in which the difference between the quantum and the classical set is the biggest one \cite{14Cabello}. This difference is precisely the one aimed by the initial state and the set of measurements used to test the violation of the KCBS inequality.

\section{$N$-cycle NC inequalities and results}
\label{sec:ngonDetails}

\begin{table}[t]
	\centering
	\caption{Experimental results for $N$-cycle witness measurements. The contextual fraction is calculated according to Eq.~(\iftoggle{arXiv}{\ref{eq:CF}}{\ref{m-eq:CF}}). Numbers in brackets are the standard errors in the means, calculated as described in App.~\ref{sec:dataAnalysis}, with each pair of observables measured 10,000 times.}
	\label{tab:gonsResults}
	\begin{tabular}{l|l|l|l|l|l}
		$N$ & $S_N$    & $S_N^{\text{(ext)}}$ & $\bar{S}_N^\text{NC}$ & $\bar{S}_5^\text{QM}$ & $\text{CF}_N$ \\ \hline
		5   & -3.926(14)   & -3.905(34)           & -3                    & -3.944                & 0.463(7)                        \\
		7   & -6.208(12)   & -6.124(39)           & -6                    & -6.271                & 0.604(6)                            \\
		11  & -10.452(10)  & -10.304(48)          & -9                    & -10.545               & 0.726(5)                           \\
		17  & -16.538(10)  & -16.538(59)          & -15                   & -16.708               & 0.769(5)                           \\
		23  & -22.530(10)  & -22.138(69)          & -21                   & -22.785               & 0.765(5)                            \\
		31  & -30.599(9)   & -30.172(79)          & -29                   & -30.840               & 0.800(4)                            \\
		41  & -40.439(11)  & -39.983(91)          & -39                   & -40.879               & 0.719(5)                           \\
		51  & -50.422(11)  & -49.740(102)         & -49                   & -50.903               & 0.711(5)                          \\
		61  & -60.279(11)  & -59.494(111)         & -59                   & -60.919               & 0.640(6)                           \\
		81  & -79.972(14)  & -79.058(128)         & -79                   & -80.939               & 0.486(7)                           \\
		101 & -99.544(17)  & -98.437(143)         & -99                   & -100.951              & 0.272(8)                          \\
		121 & -117.686(25) & -116.443(157)        & -119                  & -120.959              & -0.657(12)                       
	\end{tabular}
\end{table}

Our experimental results are summarized in Tab.~\ref{tab:gonsResults}. In order to penalize for incompatibility we again follow the scheme introduced in Ref.~\cite{15Kujala}. Accordingly, the witness for an extended $N$-cycle NC inequality is given by
\begin{equation}
\label{eq:extNcycle}
S_N^\text{(ext)} =\sum_{i=1}^{N}\av{A_{i}^{(1)}A_{i+1}^{(2)}} + \sum_{i=1}^{N}\epsilon_i \geq \bar{S}_N^\text{NC} = -N+2,
\end{equation}
where $\epsilon_i=\abs{\av{A_{i}^{(1)}}-\av{A_{i}^{(2)}}}$, as in Eq.~(\iftoggle{arXiv}{\ref{eq:extKCBS}}{\ref{m-eq:extKCBS}}). Like $S_5^\text{(ext)}$, $S_N^\text{(ext)}$ reduces to $S_N$ when $\theta = \theta_N$, and miscalibration of $\theta$ always results in $S_N^\text{(ext)} > \bar{S}_N^\text{QM}$.

\section{Comparison with previous KCBS tests}
\label{sec:KCBScomparison}

\begin{table*}[t]
	\centering
	\caption{Experimental results of previous KCBS tests. Comments in the last column are discussed in the text. The results marked by an asterisk represent our own analysis of the results table from \cite{13Um}. These differ significantly from the final results quoted in \cite{13Um}, which are currently revised due to errors in the original data analysis \cite{PrivateKim}.}
	\label{tab:allKCBS}
	\begin{tabular}{c|p{25mm}|p{35mm}|p{30mm}|p{40mm}}
		Reference & Platform & Saturation of QM limit \newline $(S_5-\bar{S}_5^\text{NC})/(\bar{S}_5^{\text{QM}}-\bar{S}_5^\text{NC})$ \newline & Signaling \newline  $\sum_{i=1}^{5}\epsilon_i /(\bar{S}_5^{\text{QM}}-\bar{S}_5^\text{NC})$ \newline & Comments  \\ \hline\hline
		Vienna, 2011 \cite{11Lapkiewicz}   & Photons   & 0.947(6)          & 0.08(3)                  & Detection loophole\newline Simultaneous measurements\newline Six observables \\
		\hline
		Stockholm, 2013 \cite{13Ahrens}  & Photons   & 0.53(11) (normal order) \newline 0.95(11) (reverse order)         & No data                    & Detection loophole\newline Order dependence     \\
		\hline
		
		Beijing, 2013 \cite{13Deng}& Photons  & 0.977(11) \newline 0.956(26)          & 0.267 \newline 0.291      & Detection loophole \newline Simultaneous measurements \newline Large signaling \\
		\hline
		Beijing, 2013 \cite{13Um}& Yb ion  & 0.589(24)*        & 0.119(24)*                & Six observables \newline Non-projective measurements     \\
		\hline
		Brisbane, 2016 \cite{16Jerger}  & Superconducting\newline circuits  & 0.520(1) (normal order)\newline 0.541(1) (reverse order)  & 0.379(2) \newline 0.379(2)                  & Large signaling     \\
		\hline
		This work  & Ca ion   & 0.969(14) (normal order)\newline 0.992(14) (reverse order)      & 0.054(31) \newline 0.050(31)            &  
	\end{tabular}
\end{table*}

We provide in Tab.~\ref{tab:allKCBS} a collection of experimental KCBS tests to benchmark our data. Aside from being among the closest to QM predictions, our analysis is the only one that we are aware of that systematically characterizes signaling and is in agreement with theoretical expectations. We have calculated the signaling the different experiments would have incurred wherever the data was made available by the authors (second-to-last column). The meanings of the different comments in the last column are:
\begin{itemize}
	\item \emph{Detection loophole}. Experiments with photons suffer from significant detector inefficiencies and setup losses. Consequently, the results assume that the registered events form an unbiased sample of all the events. The loophole does not apply when measurements are conducted with high fidelity.
	\item \emph{Simultaneous measurements}. When pairs of measurements are conducted simultaneously, it is difficult to establish an operational definition of an individual measurement \cite{16Jerger}. It has been argued that this so-called ``individual existence loophole'' can be addressed by performing measurements in a sequence \cite{09Cabello}.
	\item \emph{Six observables}. Certain experimental arrangements may not allow for performing the same measurement in the same way in every context. In order to circumvent this problem, the KCBS inequality is extended to a six-observable inequality. However, the validity of this approach has been challenged by some authors \cite{13Ahrens,13Lapkiewicz}.
	\item \emph{Order dependence}. When statistical errors dominate, the order of measurement of correlators should not influence the final outcome. Significant dependence on the order indicates the presence of uncontrolled systematic errors.
	\item \emph{Large signaling}. When signaling is large compared to observed violations, the experiment is far from the ideal assumption of compatible sharp measurements. This makes is difficult to relate the strength of contextual correlations measured in such experiments to established theoretical results. An indicator of this effect is that the statistical uncertainty in the measured witness (third column) is much smaller than the signaling term (fourth column).
	\item \emph{Non-projective measurements}. In Ref.~\cite{13Um} measurements are conducted in a sequence, but are not projective by design. Specifically, the post-measurement state, conditioned on photon detection, is a mixture of qutrit basis states. This precludes the implementation of sharp measurements. 
\end{itemize}

\section{Comparison with previous CHSH and chained Bell tests}
\label{sec:Bellcomparison}
As discussed in App.~\ref{sec:exclusivity}, both the CHSH scenario for a Bell test and the KCBS scenario for a non-contextuality test feature pentagonal graphs in the exclusivity formalism. To the best of our knowledge, there are two previous experiments aiming at the limits of correlations in Bell scenarios: one carried out by the Kwiat group in Illinois \cite{15Christensen}, the other in the group of Kurtsiefer in Singapore \cite{15Poh}. Both were carried out with photons. The former came close to 99\% of the QM prediction for the CHSH test (with statistical uncertainties well below 1\%), and the latter came down to 99.97(2)\% of the Tsirelson bound. Our result for KCBS brings us to 99.5(2)\% of the QM limit and is the first experimental demonstration of stronger-than-Bell correlations in a non-contextuality test closing the detection loophole (Fig.~\iftoggle{arXiv}{\ref{fig:KCBS}}{\ref{m-fig:KCBS}}).

In Ref.~\cite{15Christensen} they furthermore measured chained Bell inequalities with even numbers of observables all the way to $N=90$ and found a maximal value of the contextual fraction $CF_{36} = 0.874(1)$. Our results complement these with odd numbers of observables up to $N=121$ and we determine a maximal value of $CF_{31}=0.800(4)$. Aside from closing the detection loophole, we have thoroughly characterized our signaling (or incompatibility) by scanning the relevant experimental parameters. This has allowed us to identify optimal working conditions and place bounds on the amount of signaling present in the experiment. This could be relevant if the presence of signaling indicated by preliminary data analysis of previous photon experiments is confirmed \cite{18Smania}.

\section{Data repository}

The complete raw dataset is publicly available from an open repository on \url{http://www.tiqi.ethz.ch/publications-and-awards/public-datasets.html}. We encourage readers who want to expand our work with further data analysis or representations to do so.




\begin{thebibliography}{59}%
\makeatletter
\providecommand \@ifxundefined [1]{%
 \@ifx{#1\undefined}
}%
\providecommand \@ifnum [1]{%
 \ifnum #1\expandafter \@firstoftwo
 \else \expandafter \@secondoftwo
 \fi
}%
\providecommand \@ifx [1]{%
 \ifx #1\expandafter \@firstoftwo
 \else \expandafter \@secondoftwo
 \fi
}%
\providecommand \natexlab [1]{#1}%
\providecommand \enquote  [1]{``#1''}%
\providecommand \bibnamefont  [1]{#1}%
\providecommand \bibfnamefont [1]{#1}%
\providecommand \citenamefont [1]{#1}%
\providecommand \href@noop [0]{\@secondoftwo}%
\providecommand \href [0]{\begingroup \@sanitize@url \@href}%
\providecommand \@href[1]{\@@startlink{#1}\@@href}%
\providecommand \@@href[1]{\endgroup#1\@@endlink}%
\providecommand \@sanitize@url [0]{\catcode `\\12\catcode `\$12\catcode
  `\&12\catcode `\#12\catcode `\^12\catcode `\_12\catcode `\%12\relax}%
\providecommand \@@startlink[1]{}%
\providecommand \@@endlink[0]{}%
\providecommand \url  [0]{\begingroup\@sanitize@url \@url }%
\providecommand \@url [1]{\endgroup\@href {#1}{\urlprefix }}%
\providecommand \urlprefix  [0]{URL }%
\providecommand \Eprint [0]{\href }%
\providecommand \doibase [0]{http://dx.doi.org/}%
\providecommand \selectlanguage [0]{\@gobble}%
\providecommand \bibinfo  [0]{\@secondoftwo}%
\providecommand \bibfield  [0]{\@secondoftwo}%
\providecommand \translation [1]{[#1]}%
\providecommand \BibitemOpen [0]{}%
\providecommand \bibitemStop [0]{}%
\providecommand \bibitemNoStop [0]{.\EOS\space}%
\providecommand \EOS [0]{\spacefactor3000\relax}%
\providecommand \BibitemShut  [1]{\csname bibitem#1\endcsname}%
\let\auto@bib@innerbib\@empty
\bibitem [{\citenamefont {Kochen}\ and\ \citenamefont
  {Specker}(1967)}]{67Kochen}%
  \BibitemOpen
  \bibfield  {author} {\bibinfo {author} {\bibfnamefont {S.}~\bibnamefont
  {Kochen}}\ and\ \bibinfo {author} {\bibfnamefont {E.~P.}\ \bibnamefont
  {Specker}},\ }\href@noop {} {\bibfield  {journal} {\bibinfo  {journal}
  {Journal of Mathematics and Mechanics}\ }\textbf {\bibinfo {volume} {17}},\
  \bibinfo {pages} {59} (\bibinfo {year} {1967})}\BibitemShut {NoStop}%
\bibitem [{\citenamefont {Klyachko}\ \emph {et~al.}(2008)\citenamefont
  {Klyachko}, \citenamefont {Can}, \citenamefont {Binicio{\u{g}}lu},\ and\
  \citenamefont {Shumovsky}}]{08Klyachko}%
  \BibitemOpen
  \bibfield  {author} {\bibinfo {author} {\bibfnamefont {A.~A.}\ \bibnamefont
  {Klyachko}}, \bibinfo {author} {\bibfnamefont {M.~A.}\ \bibnamefont {Can}},
  \bibinfo {author} {\bibfnamefont {S.}~\bibnamefont {Binicio{\u{g}}lu}}, \
  and\ \bibinfo {author} {\bibfnamefont {A.~S.}\ \bibnamefont {Shumovsky}},\
  }\href@noop {} {\bibfield  {journal} {\bibinfo  {journal} {Physical Review
  Letters}\ }\textbf {\bibinfo {volume} {101}},\ \bibinfo {pages} {020403}
  (\bibinfo {year} {2008})}\BibitemShut {NoStop}%
\bibitem [{\citenamefont {Cabello}\ \emph {et~al.}(2014)\citenamefont
  {Cabello}, \citenamefont {Severini},\ and\ \citenamefont
  {Winter}}]{14Cabello}%
  \BibitemOpen
  \bibfield  {author} {\bibinfo {author} {\bibfnamefont {A.}~\bibnamefont
  {Cabello}}, \bibinfo {author} {\bibfnamefont {S.}~\bibnamefont {Severini}}, \
  and\ \bibinfo {author} {\bibfnamefont {A.}~\bibnamefont {Winter}},\
  }\href@noop {} {\bibfield  {journal} {\bibinfo  {journal} {Physical Review
  Letters}\ }\textbf {\bibinfo {volume} {112}},\ \bibinfo {pages} {040401}
  (\bibinfo {year} {2014})}\BibitemShut {NoStop}%
\bibitem [{\citenamefont {Heywood}\ and\ \citenamefont
  {Redhead}(1983)}]{83Heywood}%
  \BibitemOpen
  \bibfield  {author} {\bibinfo {author} {\bibfnamefont {P.}~\bibnamefont
  {Heywood}}\ and\ \bibinfo {author} {\bibfnamefont {M.~L.~G.}\ \bibnamefont
  {Redhead}},\ }\href {\doibase 10.1007/BF00729511} {\bibfield  {journal}
  {\bibinfo  {journal} {Foundations of Physics}\ }\textbf {\bibinfo {volume}
  {13}},\ \bibinfo {pages} {481} (\bibinfo {year} {1983})}\BibitemShut
  {NoStop}%
\bibitem [{\citenamefont {Cabello}\ \emph {et~al.}(2010)\citenamefont
  {Cabello}, \citenamefont {Severini},\ and\ \citenamefont
  {Winter}}]{10Cabello}%
  \BibitemOpen
  \bibfield  {author} {\bibinfo {author} {\bibfnamefont {A.}~\bibnamefont
  {Cabello}}, \bibinfo {author} {\bibfnamefont {S.}~\bibnamefont {Severini}}, \
  and\ \bibinfo {author} {\bibfnamefont {A.}~\bibnamefont {Winter}},\
  }\href@noop {} {\bibfield  {journal} {\bibinfo  {journal} {arXiv preprint
  arXiv:1010.2163}\ } (\bibinfo {year} {2010})}\BibitemShut {NoStop}%
\bibitem [{\citenamefont {Poh}\ \emph {et~al.}(2015)\citenamefont {Poh},
  \citenamefont {Joshi}, \citenamefont {Cer{\`e}}, \citenamefont {Cabello},\
  and\ \citenamefont {Kurtsiefer}}]{15Poh}%
  \BibitemOpen
  \bibfield  {author} {\bibinfo {author} {\bibfnamefont {H.~S.}\ \bibnamefont
  {Poh}}, \bibinfo {author} {\bibfnamefont {S.~K.}\ \bibnamefont {Joshi}},
  \bibinfo {author} {\bibfnamefont {A.}~\bibnamefont {Cer{\`e}}}, \bibinfo
  {author} {\bibfnamefont {A.}~\bibnamefont {Cabello}}, \ and\ \bibinfo
  {author} {\bibfnamefont {C.}~\bibnamefont {Kurtsiefer}},\ }\href@noop {}
  {\bibfield  {journal} {\bibinfo  {journal} {Physical Review Letters}\
  }\textbf {\bibinfo {volume} {115}},\ \bibinfo {pages} {180408} (\bibinfo
  {year} {2015})}\BibitemShut {NoStop}%
\bibitem [{\citenamefont {Christensen}\ \emph {et~al.}(2015)\citenamefont
  {Christensen}, \citenamefont {Liang}, \citenamefont {Brunner}, \citenamefont
  {Gisin},\ and\ \citenamefont {Kwiat}}]{15Christensen}%
  \BibitemOpen
  \bibfield  {author} {\bibinfo {author} {\bibfnamefont {B.~G.}\ \bibnamefont
  {Christensen}}, \bibinfo {author} {\bibfnamefont {Y.-C.}\ \bibnamefont
  {Liang}}, \bibinfo {author} {\bibfnamefont {N.}~\bibnamefont {Brunner}},
  \bibinfo {author} {\bibfnamefont {N.}~\bibnamefont {Gisin}}, \ and\ \bibinfo
  {author} {\bibfnamefont {P.~G.}\ \bibnamefont {Kwiat}},\ }\href {\doibase
  10.1103/PhysRevX.5.041052} {\bibfield  {journal} {\bibinfo  {journal} {Phys.
  Rev. X}\ }\textbf {\bibinfo {volume} {5}},\ \bibinfo {pages} {041052}
  (\bibinfo {year} {2015})}\BibitemShut {NoStop}%
\bibitem [{\citenamefont {Smania}\ \emph {et~al.}(2018)\citenamefont {Smania},
  \citenamefont {Kleinmann}, \citenamefont {Cabello},\ and\ \citenamefont
  {Bourennane}}]{18Smania}%
  \BibitemOpen
  \bibfield  {author} {\bibinfo {author} {\bibfnamefont {M.}~\bibnamefont
  {Smania}}, \bibinfo {author} {\bibfnamefont {M.}~\bibnamefont {Kleinmann}},
  \bibinfo {author} {\bibfnamefont {A.}~\bibnamefont {Cabello}}, \ and\
  \bibinfo {author} {\bibfnamefont {M.}~\bibnamefont {Bourennane}},\ }\href
  {http://arxiv.org/abs/1801.05739} {\bibfield  {journal} {\bibinfo  {journal}
  {arXiv}\ } (\bibinfo {year} {2018})},\ \Eprint
  {http://arxiv.org/abs/1801.05739} {1801.05739} \BibitemShut {NoStop}%
\bibitem [{\citenamefont {Lapkiewicz}\ \emph {et~al.}(2011)\citenamefont
  {Lapkiewicz}, \citenamefont {Li}, \citenamefont {Schaeff}, \citenamefont
  {Langford}, \citenamefont {Ramelow}, \citenamefont {Wie{\'s}niak},\ and\
  \citenamefont {Zeilinger}}]{11Lapkiewicz}%
  \BibitemOpen
  \bibfield  {author} {\bibinfo {author} {\bibfnamefont {R.}~\bibnamefont
  {Lapkiewicz}}, \bibinfo {author} {\bibfnamefont {P.}~\bibnamefont {Li}},
  \bibinfo {author} {\bibfnamefont {C.}~\bibnamefont {Schaeff}}, \bibinfo
  {author} {\bibfnamefont {N.~K.}\ \bibnamefont {Langford}}, \bibinfo {author}
  {\bibfnamefont {S.}~\bibnamefont {Ramelow}}, \bibinfo {author} {\bibfnamefont
  {M.}~\bibnamefont {Wie{\'s}niak}}, \ and\ \bibinfo {author} {\bibfnamefont
  {A.}~\bibnamefont {Zeilinger}},\ }\href@noop {} {\bibfield  {journal}
  {\bibinfo  {journal} {Nature}\ }\textbf {\bibinfo {volume} {474}},\ \bibinfo
  {pages} {490} (\bibinfo {year} {2011})}\BibitemShut {NoStop}%
\bibitem [{\citenamefont {Ahrens}\ \emph {et~al.}(2013)\citenamefont {Ahrens},
  \citenamefont {Amselem}, \citenamefont {Cabello},\ and\ \citenamefont
  {Bourennane}}]{13Ahrens}%
  \BibitemOpen
  \bibfield  {author} {\bibinfo {author} {\bibfnamefont {J.}~\bibnamefont
  {Ahrens}}, \bibinfo {author} {\bibfnamefont {E.}~\bibnamefont {Amselem}},
  \bibinfo {author} {\bibfnamefont {A.}~\bibnamefont {Cabello}}, \ and\
  \bibinfo {author} {\bibfnamefont {M.}~\bibnamefont {Bourennane}},\
  }\href@noop {} {\bibfield  {journal} {\bibinfo  {journal} {Scientific
  Reports}\ }\textbf {\bibinfo {volume} {3}} (\bibinfo {year}
  {2013})}\BibitemShut {NoStop}%
\bibitem [{\citenamefont {Lapkiewicz}\ \emph {et~al.}(2013)\citenamefont
  {Lapkiewicz}, \citenamefont {Li}, \citenamefont {Schaeff}, \citenamefont
  {Langford}, \citenamefont {Ramelow}, \citenamefont {Wiesniak},\ and\
  \citenamefont {Zeilinger}}]{13Lapkiewicz}%
  \BibitemOpen
  \bibfield  {author} {\bibinfo {author} {\bibfnamefont {R.}~\bibnamefont
  {Lapkiewicz}}, \bibinfo {author} {\bibfnamefont {P.}~\bibnamefont {Li}},
  \bibinfo {author} {\bibfnamefont {C.}~\bibnamefont {Schaeff}}, \bibinfo
  {author} {\bibfnamefont {N.}~\bibnamefont {Langford}}, \bibinfo {author}
  {\bibfnamefont {S.}~\bibnamefont {Ramelow}}, \bibinfo {author} {\bibfnamefont
  {M.}~\bibnamefont {Wiesniak}}, \ and\ \bibinfo {author} {\bibfnamefont
  {A.}~\bibnamefont {Zeilinger}},\ }\href {http://arxiv.org/abs/1305.5529}
  {\bibfield  {journal} {\bibinfo  {journal} {arXiv}\ } (\bibinfo {year}
  {2013})},\ \Eprint {http://arxiv.org/abs/1305.5529} {1305.5529} \BibitemShut
  {NoStop}%
\bibitem [{\citenamefont {Deng}\ \emph {et~al.}(2013)\citenamefont {Deng},
  \citenamefont {Zu}, \citenamefont {Chang}, \citenamefont {Hou}, \citenamefont
  {Yang}, \citenamefont {Wang},\ and\ \citenamefont {Duan}}]{13Deng}%
  \BibitemOpen
  \bibfield  {author} {\bibinfo {author} {\bibfnamefont {D.-L.}\ \bibnamefont
  {Deng}}, \bibinfo {author} {\bibfnamefont {C.}~\bibnamefont {Zu}}, \bibinfo
  {author} {\bibfnamefont {X.-Y.}\ \bibnamefont {Chang}}, \bibinfo {author}
  {\bibfnamefont {P.-Y.}\ \bibnamefont {Hou}}, \bibinfo {author} {\bibfnamefont
  {H.-X.}\ \bibnamefont {Yang}}, \bibinfo {author} {\bibfnamefont {Y.-X.}\
  \bibnamefont {Wang}}, \ and\ \bibinfo {author} {\bibfnamefont {L.-M.}\
  \bibnamefont {Duan}},\ }\href@noop {} {\bibfield  {journal} {\bibinfo
  {journal} {arXiv}\ } (\bibinfo {year} {2013})},\ \Eprint
  {http://arxiv.org/abs/1301.5364} {1301.5364} \BibitemShut {NoStop}%
\bibitem [{Sup()}]{SuppMat}%
  \BibitemOpen
  \href@noop {} {}\bibinfo {note} {See Supplemental Material}\BibitemShut
  {NoStop}%
\bibitem [{\citenamefont {Spekkens}(2005)}]{05Spekkens}%
  \BibitemOpen
  \bibfield  {author} {\bibinfo {author} {\bibfnamefont {R.~W.}\ \bibnamefont
  {Spekkens}},\ }\href {\doibase 10.1103/PhysRevA.71.052108} {\bibfield
  {journal} {\bibinfo  {journal} {Phys. Rev. A}\ }\textbf {\bibinfo {volume}
  {71}},\ \bibinfo {pages} {052108} (\bibinfo {year} {2005})}\BibitemShut
  {NoStop}%
\bibitem [{\citenamefont {Kunjwal}\ and\ \citenamefont
  {Spekkens}(2017)}]{17Kunjwal}%
  \BibitemOpen
  \bibfield  {author} {\bibinfo {author} {\bibfnamefont {R.}~\bibnamefont
  {Kunjwal}}\ and\ \bibinfo {author} {\bibfnamefont {R.~W.}\ \bibnamefont
  {Spekkens}},\ }\href {http://arxiv.org/abs/1708.04793} {\bibfield  {journal}
  {\bibinfo  {journal} {arXiv}\ } (\bibinfo {year} {2017})},\ \Eprint
  {http://arxiv.org/abs/1708.04793} {1708.04793} \BibitemShut {NoStop}%
\bibitem [{\citenamefont {Kujala}\ \emph {et~al.}(2015)\citenamefont {Kujala},
  \citenamefont {Dzhafarov},\ and\ \citenamefont {Larsson}}]{15Kujala}%
  \BibitemOpen
  \bibfield  {author} {\bibinfo {author} {\bibfnamefont {J.~V.}\ \bibnamefont
  {Kujala}}, \bibinfo {author} {\bibfnamefont {E.~N.}\ \bibnamefont
  {Dzhafarov}}, \ and\ \bibinfo {author} {\bibfnamefont {J.-A.}\ \bibnamefont
  {Larsson}},\ }\href {\doibase 10.1103/PhysRevLett.115.150401} {\bibfield
  {journal} {\bibinfo  {journal} {Phys. Rev. Lett.}\ }\textbf {\bibinfo
  {volume} {115}},\ \bibinfo {pages} {150401} (\bibinfo {year}
  {2015})}\BibitemShut {NoStop}%
\bibitem [{\citenamefont {Ara\'ujo}\ \emph {et~al.}(2013)\citenamefont
  {Ara\'ujo}, \citenamefont {Quintino}, \citenamefont {Budroni}, \citenamefont
  {Cunha},\ and\ \citenamefont {Cabello}}]{13Araujo}%
  \BibitemOpen
  \bibfield  {author} {\bibinfo {author} {\bibfnamefont {M.}~\bibnamefont
  {Ara\'ujo}}, \bibinfo {author} {\bibfnamefont {M.~T.}\ \bibnamefont
  {Quintino}}, \bibinfo {author} {\bibfnamefont {C.}~\bibnamefont {Budroni}},
  \bibinfo {author} {\bibfnamefont {M.~T.}\ \bibnamefont {Cunha}}, \ and\
  \bibinfo {author} {\bibfnamefont {A.}~\bibnamefont {Cabello}},\ }\href
  {\doibase 10.1103/PhysRevA.88.022118} {\bibfield  {journal} {\bibinfo
  {journal} {Phys. Rev. A}\ }\textbf {\bibinfo {volume} {88}},\ \bibinfo
  {pages} {022118} (\bibinfo {year} {2013})}\BibitemShut {NoStop}%
\bibitem [{\citenamefont {Colbeck}\ and\ \citenamefont
  {Renner}(2011)}]{11Colbeck}%
  \BibitemOpen
  \bibfield  {author} {\bibinfo {author} {\bibfnamefont {R.}~\bibnamefont
  {Colbeck}}\ and\ \bibinfo {author} {\bibfnamefont {R.}~\bibnamefont
  {Renner}},\ }\href@noop {} {\bibfield  {journal} {\bibinfo  {journal} {Nature
  communications}\ }\textbf {\bibinfo {volume} {2}},\ \bibinfo {pages} {411}
  (\bibinfo {year} {2011})}\BibitemShut {NoStop}%
\bibitem [{\citenamefont {Dhara}\ \emph {et~al.}(2013)\citenamefont {Dhara},
  \citenamefont {Prettico},\ and\ \citenamefont {Ac\'{\i}n}}]{13Dhara}%
  \BibitemOpen
  \bibfield  {author} {\bibinfo {author} {\bibfnamefont {C.}~\bibnamefont
  {Dhara}}, \bibinfo {author} {\bibfnamefont {G.}~\bibnamefont {Prettico}}, \
  and\ \bibinfo {author} {\bibfnamefont {A.}~\bibnamefont {Ac\'{\i}n}},\ }\href
  {\doibase 10.1103/PhysRevA.88.052116} {\bibfield  {journal} {\bibinfo
  {journal} {Phys. Rev. A}\ }\textbf {\bibinfo {volume} {88}},\ \bibinfo
  {pages} {052116} (\bibinfo {year} {2013})}\BibitemShut {NoStop}%
\bibitem [{\citenamefont {Abramsky}\ \emph {et~al.}(2017)\citenamefont
  {Abramsky}, \citenamefont {Barbosa},\ and\ \citenamefont
  {Mansfield}}]{17Abramsky}%
  \BibitemOpen
  \bibfield  {author} {\bibinfo {author} {\bibfnamefont {S.}~\bibnamefont
  {Abramsky}}, \bibinfo {author} {\bibfnamefont {R.~S.}\ \bibnamefont
  {Barbosa}}, \ and\ \bibinfo {author} {\bibfnamefont {S.}~\bibnamefont
  {Mansfield}},\ }\href {\doibase 10.1103/PhysRevLett.119.050504} {\bibfield
  {journal} {\bibinfo  {journal} {Phys. Rev. Lett.}\ }\textbf {\bibinfo
  {volume} {119}},\ \bibinfo {pages} {050504} (\bibinfo {year}
  {2017})}\BibitemShut {NoStop}%
\bibitem [{\citenamefont {Tan}\ \emph {et~al.}(2017)\citenamefont {Tan},
  \citenamefont {Wan}, \citenamefont {Erickson}, \citenamefont {Bierhorst},
  \citenamefont {Kienzler}, \citenamefont {Glancy}, \citenamefont {Knill},
  \citenamefont {Leibfried},\ and\ \citenamefont {Wineland}}]{17Tan}%
  \BibitemOpen
  \bibfield  {author} {\bibinfo {author} {\bibfnamefont {T.~R.}\ \bibnamefont
  {Tan}}, \bibinfo {author} {\bibfnamefont {Y.}~\bibnamefont {Wan}}, \bibinfo
  {author} {\bibfnamefont {S.}~\bibnamefont {Erickson}}, \bibinfo {author}
  {\bibfnamefont {P.}~\bibnamefont {Bierhorst}}, \bibinfo {author}
  {\bibfnamefont {D.}~\bibnamefont {Kienzler}}, \bibinfo {author}
  {\bibfnamefont {S.}~\bibnamefont {Glancy}}, \bibinfo {author} {\bibfnamefont
  {E.}~\bibnamefont {Knill}}, \bibinfo {author} {\bibfnamefont
  {D.}~\bibnamefont {Leibfried}}, \ and\ \bibinfo {author} {\bibfnamefont
  {D.~J.}\ \bibnamefont {Wineland}},\ }\href {\doibase
  10.1103/PhysRevLett.118.130403} {\bibfield  {journal} {\bibinfo  {journal}
  {Phys. Rev. Lett.}\ }\textbf {\bibinfo {volume} {118}},\ \bibinfo {pages}
  {130403} (\bibinfo {year} {2017})}\BibitemShut {NoStop}%
\bibitem [{\citenamefont {Arias}\ \emph {et~al.}(2015)\citenamefont {Arias},
  \citenamefont {Ca\~nas}, \citenamefont {G\'omez}, \citenamefont {Barra},
  \citenamefont {Xavier}, \citenamefont {Lima}, \citenamefont {D'Ambrosio},
  \citenamefont {Baccari}, \citenamefont {Sciarrino},\ and\ \citenamefont
  {Cabello}}]{15Arias}%
  \BibitemOpen
  \bibfield  {author} {\bibinfo {author} {\bibfnamefont {M.}~\bibnamefont
  {Arias}}, \bibinfo {author} {\bibfnamefont {G.}~\bibnamefont {Ca\~nas}},
  \bibinfo {author} {\bibfnamefont {E.~S.}\ \bibnamefont {G\'omez}}, \bibinfo
  {author} {\bibfnamefont {J.~F.}\ \bibnamefont {Barra}}, \bibinfo {author}
  {\bibfnamefont {G.~B.}\ \bibnamefont {Xavier}}, \bibinfo {author}
  {\bibfnamefont {G.}~\bibnamefont {Lima}}, \bibinfo {author} {\bibfnamefont
  {V.}~\bibnamefont {D'Ambrosio}}, \bibinfo {author} {\bibfnamefont
  {F.}~\bibnamefont {Baccari}}, \bibinfo {author} {\bibfnamefont
  {F.}~\bibnamefont {Sciarrino}}, \ and\ \bibinfo {author} {\bibfnamefont
  {A.}~\bibnamefont {Cabello}},\ }\href {\doibase 10.1103/PhysRevA.92.032126}
  {\bibfield  {journal} {\bibinfo  {journal} {Phys. Rev. A}\ }\textbf {\bibinfo
  {volume} {92}},\ \bibinfo {pages} {032126} (\bibinfo {year}
  {2015})}\BibitemShut {NoStop}%
\bibitem [{\citenamefont {Cabello}(2013)}]{13Cabello}%
  \BibitemOpen
  \bibfield  {author} {\bibinfo {author} {\bibfnamefont {A.}~\bibnamefont
  {Cabello}},\ }\href {\doibase 10.1103/PhysRevLett.110.060402} {\bibfield
  {journal} {\bibinfo  {journal} {Phys. Rev. Lett.}\ }\textbf {\bibinfo
  {volume} {110}},\ \bibinfo {pages} {060402} (\bibinfo {year}
  {2013})}\BibitemShut {NoStop}%
\bibitem [{\citenamefont {Abramsky}\ and\ \citenamefont
  {Brandenburger}(2011)}]{11Abramsky}%
  \BibitemOpen
  \bibfield  {author} {\bibinfo {author} {\bibfnamefont {S.}~\bibnamefont
  {Abramsky}}\ and\ \bibinfo {author} {\bibfnamefont {A.}~\bibnamefont
  {Brandenburger}},\ }\href {http://stacks.iop.org/1367-2630/13/i=11/a=113036}
  {\bibfield  {journal} {\bibinfo  {journal} {New Journal of Physics}\ }\textbf
  {\bibinfo {volume} {13}},\ \bibinfo {pages} {113036} (\bibinfo {year}
  {2011})}\BibitemShut {NoStop}%
\bibitem [{\citenamefont {G{\"u}hne}\ \emph {et~al.}(2010)\citenamefont
  {G{\"u}hne}, \citenamefont {Kleinmann}, \citenamefont {Cabello},
  \citenamefont {Larsson}, \citenamefont {Kirchmair}, \citenamefont
  {Z{\"a}hringer}, \citenamefont {Gerritsma},\ and\ \citenamefont
  {Roos}}]{10Guhne}%
  \BibitemOpen
  \bibfield  {author} {\bibinfo {author} {\bibfnamefont {O.}~\bibnamefont
  {G{\"u}hne}}, \bibinfo {author} {\bibfnamefont {M.}~\bibnamefont
  {Kleinmann}}, \bibinfo {author} {\bibfnamefont {A.}~\bibnamefont {Cabello}},
  \bibinfo {author} {\bibfnamefont {J.-{\AA}.}\ \bibnamefont {Larsson}},
  \bibinfo {author} {\bibfnamefont {G.}~\bibnamefont {Kirchmair}}, \bibinfo
  {author} {\bibfnamefont {F.}~\bibnamefont {Z{\"a}hringer}}, \bibinfo {author}
  {\bibfnamefont {R.}~\bibnamefont {Gerritsma}}, \ and\ \bibinfo {author}
  {\bibfnamefont {C.~F.}\ \bibnamefont {Roos}},\ }\href@noop {} {\bibfield
  {journal} {\bibinfo  {journal} {Physical Review A}\ }\textbf {\bibinfo
  {volume} {81}},\ \bibinfo {pages} {022121} (\bibinfo {year}
  {2010})}\BibitemShut {NoStop}%
\bibitem [{\citenamefont {Larsson}(2014)}]{14Larsson}%
  \BibitemOpen
  \bibfield  {author} {\bibinfo {author} {\bibfnamefont {J.-{\AA}.}\
  \bibnamefont {Larsson}},\ }\href
  {http://stacks.iop.org/1751-8121/47/i=42/a=424003} {\bibfield  {journal}
  {\bibinfo  {journal} {Journal of Physics A: Mathematical and Theoretical}\
  }\textbf {\bibinfo {volume} {47}},\ \bibinfo {pages} {424003} (\bibinfo
  {year} {2014})}\BibitemShut {NoStop}%
\bibitem [{\citenamefont {Leupold}\ \emph {et~al.}(2017)\citenamefont
  {Leupold}, \citenamefont {Malinowski}, \citenamefont {Zhang}, \citenamefont
  {Cabello}, \citenamefont {Alonso},\ and\ \citenamefont {Home}}]{17Leupold}%
  \BibitemOpen
  \bibfield  {author} {\bibinfo {author} {\bibfnamefont {F.~M.}\ \bibnamefont
  {Leupold}}, \bibinfo {author} {\bibfnamefont {M.}~\bibnamefont {Malinowski}},
  \bibinfo {author} {\bibfnamefont {C.}~\bibnamefont {Zhang}}, \bibinfo
  {author} {\bibfnamefont {A.}~\bibnamefont {Cabello}}, \bibinfo {author}
  {\bibfnamefont {J.}~\bibnamefont {Alonso}}, \ and\ \bibinfo {author}
  {\bibfnamefont {J.~P.}\ \bibnamefont {Home}},\ }\href@noop {} {\bibfield
  {journal} {\bibinfo  {journal} {arXiv:1706.07370}\ } (\bibinfo {year}
  {2017})}\BibitemShut {NoStop}%
\bibitem [{\citenamefont {Alonso}\ \emph {et~al.}(2016)\citenamefont {Alonso},
  \citenamefont {Leupold}, \citenamefont {Sol\`er}, \citenamefont {Fadel},
  \citenamefont {Marinelli}, \citenamefont {Keitch}, \citenamefont
  {Negnevitsky},\ and\ \citenamefont {Home}}]{16Alonso}%
  \BibitemOpen
  \bibfield  {author} {\bibinfo {author} {\bibfnamefont {J.}~\bibnamefont
  {Alonso}}, \bibinfo {author} {\bibfnamefont {F.~M.}\ \bibnamefont {Leupold}},
  \bibinfo {author} {\bibfnamefont {Z.~U.}\ \bibnamefont {Sol\`er}}, \bibinfo
  {author} {\bibfnamefont {M.}~\bibnamefont {Fadel}}, \bibinfo {author}
  {\bibfnamefont {M.}~\bibnamefont {Marinelli}}, \bibinfo {author}
  {\bibfnamefont {B.~C.}\ \bibnamefont {Keitch}}, \bibinfo {author}
  {\bibfnamefont {V.}~\bibnamefont {Negnevitsky}}, \ and\ \bibinfo {author}
  {\bibfnamefont {J.~P.}\ \bibnamefont {Home}},\ }\href@noop {} {\bibfield
  {journal} {\bibinfo  {journal} {Nature Communications}\ }\textbf {\bibinfo
  {volume} {7}},\ \bibinfo {pages} {11243} (\bibinfo {year}
  {2016})}\BibitemShut {NoStop}%
\bibitem [{\citenamefont {Jerger}\ \emph {et~al.}(2016)\citenamefont {Jerger},
  \citenamefont {Reshitnyk}, \citenamefont {Oppliger}, \citenamefont
  {Poto{\v{c}}nik}, \citenamefont {Mondal}, \citenamefont {Wallraff},
  \citenamefont {Goodenough}, \citenamefont {Wehner}, \citenamefont
  {Juliusson}, \citenamefont {Langford} \emph {et~al.}}]{16Jerger}%
  \BibitemOpen
  \bibfield  {author} {\bibinfo {author} {\bibfnamefont {M.}~\bibnamefont
  {Jerger}}, \bibinfo {author} {\bibfnamefont {Y.}~\bibnamefont {Reshitnyk}},
  \bibinfo {author} {\bibfnamefont {M.}~\bibnamefont {Oppliger}}, \bibinfo
  {author} {\bibfnamefont {A.}~\bibnamefont {Poto{\v{c}}nik}}, \bibinfo
  {author} {\bibfnamefont {M.}~\bibnamefont {Mondal}}, \bibinfo {author}
  {\bibfnamefont {A.}~\bibnamefont {Wallraff}}, \bibinfo {author}
  {\bibfnamefont {K.}~\bibnamefont {Goodenough}}, \bibinfo {author}
  {\bibfnamefont {S.}~\bibnamefont {Wehner}}, \bibinfo {author} {\bibfnamefont
  {K.}~\bibnamefont {Juliusson}}, \bibinfo {author} {\bibfnamefont {N.~K.}\
  \bibnamefont {Langford}},  \emph {et~al.},\ }\href@noop {} {\bibfield
  {journal} {\bibinfo  {journal} {Nature communications}\ }\textbf {\bibinfo
  {volume} {7}} (\bibinfo {year} {2016})}\BibitemShut {NoStop}%
\bibitem [{\citenamefont {Chiribella}\ and\ \citenamefont
  {Yuan}(2014)}]{14Chiribella}%
  \BibitemOpen
  \bibfield  {author} {\bibinfo {author} {\bibfnamefont {G.}~\bibnamefont
  {Chiribella}}\ and\ \bibinfo {author} {\bibfnamefont {X.}~\bibnamefont
  {Yuan}},\ }\href {http://arxiv.org/abs/1404.3348v2} {\bibfield  {journal}
  {\bibinfo  {journal} {arXiv}\ } (\bibinfo {year} {2014})},\ \Eprint
  {http://arxiv.org/abs/1404.3348v2} {1404.3348v2} \BibitemShut {NoStop}%
\bibitem [{\citenamefont {Kent}(1999)}]{99Kent}%
  \BibitemOpen
  \bibfield  {author} {\bibinfo {author} {\bibfnamefont {A.}~\bibnamefont
  {Kent}},\ }\href {\doibase 10.1103/PhysRevLett.83.3755} {\bibfield  {journal}
  {\bibinfo  {journal} {Phys. Rev. Lett.}\ }\textbf {\bibinfo {volume} {83}},\
  \bibinfo {pages} {3755} (\bibinfo {year} {1999})}\BibitemShut {NoStop}%
\bibitem [{\citenamefont {Clifton}\ and\ \citenamefont
  {Kent}(2000)}]{00Clifton}%
  \BibitemOpen
  \bibfield  {author} {\bibinfo {author} {\bibfnamefont {R.}~\bibnamefont
  {Clifton}}\ and\ \bibinfo {author} {\bibfnamefont {A.}~\bibnamefont {Kent}},\
  }in\ \href@noop {} {\emph {\bibinfo {booktitle} {Proceedings of the Royal
  Society of London A: Mathematical, Physical and Engineering Sciences}}},\
  Vol.\ \bibinfo {volume} {456}\ (\bibinfo {organization} {The Royal Society},\
  \bibinfo {year} {2000})\ pp.\ \bibinfo {pages} {2101--2114}\BibitemShut
  {NoStop}%
\bibitem [{\citenamefont {Cabello}(2002)}]{02Cabello}%
  \BibitemOpen
  \bibfield  {author} {\bibinfo {author} {\bibfnamefont {A.}~\bibnamefont
  {Cabello}},\ }\href {\doibase 10.1103/PhysRevA.65.052101} {\bibfield
  {journal} {\bibinfo  {journal} {Phys. Rev. A}\ }\textbf {\bibinfo {volume}
  {65}},\ \bibinfo {pages} {052101} (\bibinfo {year} {2002})}\BibitemShut
  {NoStop}%
\bibitem [{\citenamefont {Hu}\ \emph {et~al.}(2016)\citenamefont {Hu},
  \citenamefont {Chen}, \citenamefont {Liu}, \citenamefont {Guo}, \citenamefont
  {Huang}, \citenamefont {Zhou}, \citenamefont {Han}, \citenamefont {Li},\ and\
  \citenamefont {Guo}}]{16Hu}%
  \BibitemOpen
  \bibfield  {author} {\bibinfo {author} {\bibfnamefont {X.-M.}\ \bibnamefont
  {Hu}}, \bibinfo {author} {\bibfnamefont {J.-S.}\ \bibnamefont {Chen}},
  \bibinfo {author} {\bibfnamefont {B.-H.}\ \bibnamefont {Liu}}, \bibinfo
  {author} {\bibfnamefont {Y.}~\bibnamefont {Guo}}, \bibinfo {author}
  {\bibfnamefont {Y.-F.}\ \bibnamefont {Huang}}, \bibinfo {author}
  {\bibfnamefont {Z.-Q.}\ \bibnamefont {Zhou}}, \bibinfo {author}
  {\bibfnamefont {Y.-J.}\ \bibnamefont {Han}}, \bibinfo {author} {\bibfnamefont
  {C.-F.}\ \bibnamefont {Li}}, \ and\ \bibinfo {author} {\bibfnamefont {G.-C.}\
  \bibnamefont {Guo}},\ }\href@noop {} {\bibfield  {journal} {\bibinfo
  {journal} {Physical Review Letters}\ }\textbf {\bibinfo {volume} {117}},\
  \bibinfo {pages} {170403} (\bibinfo {year} {2016})}\BibitemShut {NoStop}%
\bibitem [{\citenamefont {Cabello}(2016)}]{16Cabello2}%
  \BibitemOpen
  \bibfield  {author} {\bibinfo {author} {\bibfnamefont {A.}~\bibnamefont
  {Cabello}},\ }\href {\doibase 10.1103/PhysRevA.93.032102} {\bibfield
  {journal} {\bibinfo  {journal} {Phys. Rev. A}\ }\textbf {\bibinfo {volume}
  {93}},\ \bibinfo {pages} {032102} (\bibinfo {year} {2016})}\BibitemShut
  {NoStop}%
\bibitem [{Note1()}]{Note1}%
  \BibitemOpen
  \bibinfo {note} {The complete raw dataset is publicly available from an open
  repository on \protect \url
  {http://www.tiqi.ethz.ch/publications-and-awards/public-datasets.html}.}\BibitemShut
  {Stop}%
\bibitem [{\citenamefont {Liang}\ \emph {et~al.}(2011)\citenamefont {Liang},
  \citenamefont {Spekkens},\ and\ \citenamefont {Wiseman}}]{11Liang}%
  \BibitemOpen
  \bibfield  {author} {\bibinfo {author} {\bibfnamefont {Y.-C.}\ \bibnamefont
  {Liang}}, \bibinfo {author} {\bibfnamefont {R.~W.}\ \bibnamefont {Spekkens}},
  \ and\ \bibinfo {author} {\bibfnamefont {H.~M.}\ \bibnamefont {Wiseman}},\
  }\href {\doibase https://doi.org/10.1016/j.physrep.2011.05.001} {\bibfield
  {journal} {\bibinfo  {journal} {Physics Reports}\ }\textbf {\bibinfo {volume}
  {506}},\ \bibinfo {pages} {1 } (\bibinfo {year} {2011})}\BibitemShut
  {NoStop}%
\bibitem [{\citenamefont {Sadiq}\ \emph {et~al.}(2013)\citenamefont {Sadiq},
  \citenamefont {Badzi{\k{a}}g}, \citenamefont {Bourennane},\ and\
  \citenamefont {Cabello}}]{13Sadiq}%
  \BibitemOpen
  \bibfield  {author} {\bibinfo {author} {\bibfnamefont {M.}~\bibnamefont
  {Sadiq}}, \bibinfo {author} {\bibfnamefont {P.}~\bibnamefont
  {Badzi{\k{a}}g}}, \bibinfo {author} {\bibfnamefont {M.}~\bibnamefont
  {Bourennane}}, \ and\ \bibinfo {author} {\bibfnamefont {A.}~\bibnamefont
  {Cabello}},\ }\href@noop {} {\bibfield  {journal} {\bibinfo  {journal}
  {Physical Review A}\ }\textbf {\bibinfo {volume} {87}},\ \bibinfo {pages}
  {012128} (\bibinfo {year} {2013})}\BibitemShut {NoStop}%
\bibitem [{\citenamefont {Amselem}\ \emph {et~al.}(2012)\citenamefont
  {Amselem}, \citenamefont {Danielsen}, \citenamefont {L\'opez-Tarrida},
  \citenamefont {Portillo}, \citenamefont {Bourennane},\ and\ \citenamefont
  {Cabello}}]{12Amselem}%
  \BibitemOpen
  \bibfield  {author} {\bibinfo {author} {\bibfnamefont {E.}~\bibnamefont
  {Amselem}}, \bibinfo {author} {\bibfnamefont {L.~E.}\ \bibnamefont
  {Danielsen}}, \bibinfo {author} {\bibfnamefont {A.~J.}\ \bibnamefont
  {L\'opez-Tarrida}}, \bibinfo {author} {\bibfnamefont {J.~R.}\ \bibnamefont
  {Portillo}}, \bibinfo {author} {\bibfnamefont {M.}~\bibnamefont
  {Bourennane}}, \ and\ \bibinfo {author} {\bibfnamefont {A.}~\bibnamefont
  {Cabello}},\ }\href {\doibase 10.1103/PhysRevLett.108.200405} {\bibfield
  {journal} {\bibinfo  {journal} {Phys. Rev. Lett.}\ }\textbf {\bibinfo
  {volume} {108}},\ \bibinfo {pages} {200405} (\bibinfo {year}
  {2012})}\BibitemShut {NoStop}%
\bibitem [{\citenamefont {Kofler}\ and\ \citenamefont
  {Brukner}(2007)}]{07Kofler}%
  \BibitemOpen
  \bibfield  {author} {\bibinfo {author} {\bibfnamefont {J.}~\bibnamefont
  {Kofler}}\ and\ \bibinfo {author} {\bibfnamefont {C.}~\bibnamefont
  {Brukner}},\ }\href {\doibase 10.1103/PhysRevLett.99.180403} {\bibfield
  {journal} {\bibinfo  {journal} {Phys. Rev. Lett.}\ }\textbf {\bibinfo
  {volume} {99}},\ \bibinfo {pages} {180403} (\bibinfo {year}
  {2007})}\BibitemShut {NoStop}%
\bibitem [{\citenamefont {Jeong}\ \emph {et~al.}(2014)\citenamefont {Jeong},
  \citenamefont {Lim},\ and\ \citenamefont {Kim}}]{14Jeong}%
  \BibitemOpen
  \bibfield  {author} {\bibinfo {author} {\bibfnamefont {H.}~\bibnamefont
  {Jeong}}, \bibinfo {author} {\bibfnamefont {Y.}~\bibnamefont {Lim}}, \ and\
  \bibinfo {author} {\bibfnamefont {M.~S.}\ \bibnamefont {Kim}},\ }\href
  {\doibase 10.1103/PhysRevLett.112.010402} {\bibfield  {journal} {\bibinfo
  {journal} {Phys. Rev. Lett.}\ }\textbf {\bibinfo {volume} {112}},\ \bibinfo
  {pages} {010402} (\bibinfo {year} {2014})}\BibitemShut {NoStop}%
\bibitem [{\citenamefont {Cabello}\ and\ \citenamefont
  {Cunha}(2011)}]{11Cabello}%
  \BibitemOpen
  \bibfield  {author} {\bibinfo {author} {\bibfnamefont {A.}~\bibnamefont
  {Cabello}}\ and\ \bibinfo {author} {\bibfnamefont {M.~T.}\ \bibnamefont
  {Cunha}},\ }\href@noop {} {\bibfield  {journal} {\bibinfo  {journal}
  {Physical Review Letters}\ }\textbf {\bibinfo {volume} {106}},\ \bibinfo
  {pages} {190401} (\bibinfo {year} {2011})}\BibitemShut {NoStop}%
\bibitem [{\citenamefont {Zhan}\ \emph {et~al.}(2016)\citenamefont {Zhan},
  \citenamefont {Zhang}, \citenamefont {Li}, \citenamefont {Zhang},
  \citenamefont {Sanders},\ and\ \citenamefont {Xue}}]{16Zhan}%
  \BibitemOpen
  \bibfield  {author} {\bibinfo {author} {\bibfnamefont {X.}~\bibnamefont
  {Zhan}}, \bibinfo {author} {\bibfnamefont {X.}~\bibnamefont {Zhang}},
  \bibinfo {author} {\bibfnamefont {J.}~\bibnamefont {Li}}, \bibinfo {author}
  {\bibfnamefont {Y.}~\bibnamefont {Zhang}}, \bibinfo {author} {\bibfnamefont
  {B.~C.}\ \bibnamefont {Sanders}}, \ and\ \bibinfo {author} {\bibfnamefont
  {P.}~\bibnamefont {Xue}},\ }\href@noop {} {\bibfield  {journal} {\bibinfo
  {journal} {Physical Review Letters}\ }\textbf {\bibinfo {volume} {116}},\
  \bibinfo {pages} {090401} (\bibinfo {year} {2016})}\BibitemShut {NoStop}%
\bibitem [{\citenamefont {Mazurek}\ \emph {et~al.}(2016)\citenamefont
  {Mazurek}, \citenamefont {Pusey}, \citenamefont {Kunjwal}, \citenamefont
  {Resch},\ and\ \citenamefont {Spekkens}}]{16Mazurek}%
  \BibitemOpen
  \bibfield  {author} {\bibinfo {author} {\bibfnamefont {M.~D.}\ \bibnamefont
  {Mazurek}}, \bibinfo {author} {\bibfnamefont {M.~F.}\ \bibnamefont {Pusey}},
  \bibinfo {author} {\bibfnamefont {R.}~\bibnamefont {Kunjwal}}, \bibinfo
  {author} {\bibfnamefont {K.~J.}\ \bibnamefont {Resch}}, \ and\ \bibinfo
  {author} {\bibfnamefont {R.~W.}\ \bibnamefont {Spekkens}},\ }\href@noop {}
  {\bibfield  {journal} {\bibinfo  {journal} {Nature communications}\ }\textbf
  {\bibinfo {volume} {7}} (\bibinfo {year} {2016})}\BibitemShut {NoStop}%
\bibitem [{\citenamefont {Wineland}\ \emph {et~al.}(1998)\citenamefont
  {Wineland}, \citenamefont {Monroe}, \citenamefont {Itano}, \citenamefont
  {Leibfried}, \citenamefont {King},\ and\ \citenamefont
  {Meekhof}}]{98Wineland2}%
  \BibitemOpen
  \bibfield  {author} {\bibinfo {author} {\bibfnamefont {D.~J.}\ \bibnamefont
  {Wineland}}, \bibinfo {author} {\bibfnamefont {C.}~\bibnamefont {Monroe}},
  \bibinfo {author} {\bibfnamefont {W.~M.}\ \bibnamefont {Itano}}, \bibinfo
  {author} {\bibfnamefont {D.}~\bibnamefont {Leibfried}}, \bibinfo {author}
  {\bibfnamefont {B.~E.}\ \bibnamefont {King}}, \ and\ \bibinfo {author}
  {\bibfnamefont {D.~M.}\ \bibnamefont {Meekhof}},\ }\href@noop {} {\bibfield
  {journal} {\bibinfo  {journal} {J. Res. Natl. Inst. Stand. Technol.}\
  }\textbf {\bibinfo {volume} {103}},\ \bibinfo {pages} {259} (\bibinfo {year}
  {1998})}\BibitemShut {NoStop}%
\bibitem [{\citenamefont {L{\"u}ders}(1950)}]{50Luders}%
  \BibitemOpen
  \bibfield  {author} {\bibinfo {author} {\bibfnamefont {G.}~\bibnamefont
  {L{\"u}ders}},\ }\href@noop {} {\bibfield  {journal} {\bibinfo  {journal}
  {Annalen der Physik}\ }\textbf {\bibinfo {volume} {443}},\ \bibinfo {pages}
  {322} (\bibinfo {year} {1950})}\BibitemShut {NoStop}%
\bibitem [{\citenamefont {Leone}\ \emph {et~al.}(1961)\citenamefont {Leone},
  \citenamefont {Nelson},\ and\ \citenamefont {Nottingham}}]{61Leone}%
  \BibitemOpen
  \bibfield  {author} {\bibinfo {author} {\bibfnamefont {F.~C.}\ \bibnamefont
  {Leone}}, \bibinfo {author} {\bibfnamefont {L.~S.}\ \bibnamefont {Nelson}}, \
  and\ \bibinfo {author} {\bibfnamefont {R.~B.}\ \bibnamefont {Nottingham}},\
  }\href {\doibase 10.1080/00401706.1961.10489974} {\bibfield  {journal}
  {\bibinfo  {journal} {Technometrics}\ }\textbf {\bibinfo {volume} {3}},\
  \bibinfo {pages} {543} (\bibinfo {year} {1961})}\BibitemShut {NoStop}%
\bibitem [{\citenamefont {Bell}(1966)}]{66Bell}%
  \BibitemOpen
  \bibfield  {author} {\bibinfo {author} {\bibfnamefont {J.~S.}\ \bibnamefont
  {Bell}},\ }\href {\doibase 10.1103/RevModPhys.38.447} {\bibfield  {journal}
  {\bibinfo  {journal} {Rev. Mod. Phys.}\ }\textbf {\bibinfo {volume} {38}},\
  \bibinfo {pages} {447} (\bibinfo {year} {1966})}\BibitemShut {NoStop}%
\bibitem [{\citenamefont {Pearle}(1970)}]{70Pearle}%
  \BibitemOpen
  \bibfield  {author} {\bibinfo {author} {\bibfnamefont {P.~M.}\ \bibnamefont
  {Pearle}},\ }\href {\doibase 10.1103/PhysRevD.2.1418} {\bibfield  {journal}
  {\bibinfo  {journal} {Phys. Rev. D}\ }\textbf {\bibinfo {volume} {2}},\
  \bibinfo {pages} {1418} (\bibinfo {year} {1970})}\BibitemShut {NoStop}%
\bibitem [{\citenamefont {Braunstein}\ and\ \citenamefont
  {Caves}(1990)}]{90Braunstein}%
  \BibitemOpen
  \bibfield  {author} {\bibinfo {author} {\bibfnamefont {S.~L.}\ \bibnamefont
  {Braunstein}}\ and\ \bibinfo {author} {\bibfnamefont {C.~M.}\ \bibnamefont
  {Caves}},\ }\href {\doibase https://doi.org/10.1016/0003-4916(90)90339-P}
  {\bibfield  {journal} {\bibinfo  {journal} {Annals of Physics}\ }\textbf
  {\bibinfo {volume} {202}},\ \bibinfo {pages} {22 } (\bibinfo {year}
  {1990})}\BibitemShut {NoStop}%
\bibitem [{\citenamefont {Fine}(1982)}]{82Fine}%
  \BibitemOpen
  \bibfield  {author} {\bibinfo {author} {\bibfnamefont {A.}~\bibnamefont
  {Fine}},\ }\href {\doibase 10.1103/PhysRevLett.48.291} {\bibfield  {journal}
  {\bibinfo  {journal} {Phys. Rev. Lett.}\ }\textbf {\bibinfo {volume} {48}},\
  \bibinfo {pages} {291} (\bibinfo {year} {1982})}\BibitemShut {NoStop}%
\bibitem [{\citenamefont {Sliwa}(2003)}]{03Sliwa}%
  \BibitemOpen
  \bibfield  {author} {\bibinfo {author} {\bibfnamefont {C.}~\bibnamefont
  {Sliwa}},\ }\href {\doibase https://doi.org/10.1016/S0375-9601(03)01115-0}
  {\bibfield  {journal} {\bibinfo  {journal} {Physics Letters A}\ }\textbf
  {\bibinfo {volume} {317}},\ \bibinfo {pages} {165 } (\bibinfo {year}
  {2003})}\BibitemShut {NoStop}%
\bibitem [{\citenamefont {Collins}\ and\ \citenamefont
  {Gisin}(2004)}]{04Collins}%
  \BibitemOpen
  \bibfield  {author} {\bibinfo {author} {\bibfnamefont {D.}~\bibnamefont
  {Collins}}\ and\ \bibinfo {author} {\bibfnamefont {N.}~\bibnamefont
  {Gisin}},\ }\href {http://stacks.iop.org/0305-4470/37/i=5/a=021} {\bibfield
  {journal} {\bibinfo  {journal} {Journal of Physics A: Mathematical and
  General}\ }\textbf {\bibinfo {volume} {37}},\ \bibinfo {pages} {1775}
  (\bibinfo {year} {2004})}\BibitemShut {NoStop}%
\bibitem [{\citenamefont {L\'opez-Rosa}\ \emph {et~al.}(2016)\citenamefont
  {L\'opez-Rosa}, \citenamefont {Xu},\ and\ \citenamefont {Cabello}}]{16Lopez}%
  \BibitemOpen
  \bibfield  {author} {\bibinfo {author} {\bibfnamefont {S.}~\bibnamefont
  {L\'opez-Rosa}}, \bibinfo {author} {\bibfnamefont {Z.-P.}\ \bibnamefont
  {Xu}}, \ and\ \bibinfo {author} {\bibfnamefont {A.}~\bibnamefont {Cabello}},\
  }\href {\doibase 10.1103/PhysRevA.94.062121} {\bibfield  {journal} {\bibinfo
  {journal} {Phys. Rev. A}\ }\textbf {\bibinfo {volume} {94}},\ \bibinfo
  {pages} {062121} (\bibinfo {year} {2016})}\BibitemShut {NoStop}%
\bibitem [{\citenamefont {Rabelo}\ \emph {et~al.}(2017)\citenamefont {Rabelo},
  \citenamefont {Cunha},\ and\ \citenamefont {Cabello}}]{17Rabelo}%
  \BibitemOpen
  \bibfield  {author} {\bibinfo {author} {\bibfnamefont {R.}~\bibnamefont
  {Rabelo}}, \bibinfo {author} {\bibfnamefont {M.~T.}\ \bibnamefont {Cunha}}, \
  and\ \bibinfo {author} {\bibfnamefont {A.}~\bibnamefont {Cabello}},\
  }\href@noop {} {\enquote {\bibinfo {title} {In preparation},}\ } (\bibinfo
  {year} {2017})\BibitemShut {NoStop}%
\bibitem [{\citenamefont {Clauser}\ \emph {et~al.}(1969)\citenamefont
  {Clauser}, \citenamefont {Horne}, \citenamefont {Shimony},\ and\
  \citenamefont {Holt}}]{69Clauser}%
  \BibitemOpen
  \bibfield  {author} {\bibinfo {author} {\bibfnamefont {J.~F.}\ \bibnamefont
  {Clauser}}, \bibinfo {author} {\bibfnamefont {M.~A.}\ \bibnamefont {Horne}},
  \bibinfo {author} {\bibfnamefont {A.}~\bibnamefont {Shimony}}, \ and\
  \bibinfo {author} {\bibfnamefont {R.~A.}\ \bibnamefont {Holt}},\ }\href@noop
  {} {\bibfield  {journal} {\bibinfo  {journal} {Physical Review Letters}\
  }\textbf {\bibinfo {volume} {23}},\ \bibinfo {pages} {880} (\bibinfo {year}
  {1969})}\BibitemShut {NoStop}%
\bibitem [{\citenamefont {Um}\ \emph {et~al.}(2013)\citenamefont {Um},
  \citenamefont {Zhang}, \citenamefont {Zhang}, \citenamefont {Wang},
  \citenamefont {Yangchao}, \citenamefont {Deng}, \citenamefont {Duan},\ and\
  \citenamefont {Kim}}]{13Um}%
  \BibitemOpen
  \bibfield  {author} {\bibinfo {author} {\bibfnamefont {M.}~\bibnamefont
  {Um}}, \bibinfo {author} {\bibfnamefont {X.}~\bibnamefont {Zhang}}, \bibinfo
  {author} {\bibfnamefont {J.}~\bibnamefont {Zhang}}, \bibinfo {author}
  {\bibfnamefont {Y.}~\bibnamefont {Wang}}, \bibinfo {author} {\bibfnamefont
  {S.}~\bibnamefont {Yangchao}}, \bibinfo {author} {\bibfnamefont {D.-L.}\
  \bibnamefont {Deng}}, \bibinfo {author} {\bibfnamefont {L.-M.}\ \bibnamefont
  {Duan}}, \ and\ \bibinfo {author} {\bibfnamefont {K.}~\bibnamefont {Kim}},\
  }\href@noop {} {\bibfield  {journal} {\bibinfo  {journal} {Scientific
  Reports}\ }\textbf {\bibinfo {volume} {3}} (\bibinfo {year}
  {2013})}\BibitemShut {NoStop}%
\bibitem [{\citenamefont {Kim}(2017)}]{PrivateKim}%
  \BibitemOpen
  \bibfield  {author} {\bibinfo {author} {\bibfnamefont {K.}~\bibnamefont
  {Kim}},\ }\href@noop {} {\enquote {\bibinfo {title} {Private
  communication},}\ } (\bibinfo {year} {2017})\BibitemShut {NoStop}%
\bibitem [{\citenamefont {Cabello}(2009)}]{09Cabello}%
  \BibitemOpen
  \bibfield  {author} {\bibinfo {author} {\bibfnamefont {A.}~\bibnamefont
  {Cabello}},\ }in\ \href@noop {} {\emph {\bibinfo {booktitle} {FOUNDATIONS OF
  PROBABILITY AND PHYSICS 5}}},\ Vol.\ \bibinfo {volume} {1101}\ (\bibinfo
  {organization} {AIP Publishing},\ \bibinfo {year} {2009})\ pp.\ \bibinfo
  {pages} {246--254}\BibitemShut {NoStop}%
\end{thebibliography}
%

\end{document}